%% file: ScalarLQPaper.tex
\documentclass[cernpreprint,texlive=2011,txfonts,UKenglish]{atlasdoc}

\usepackage{multirow}
\usepackage{ulem}
\usepackage{subfig}
\usepackage{lscape}
\usepackage{verbatim}
\usepackage{float}
\usepackage{graphicx}
\usepackage{url}
\usepackage{caption}
\usepackage{hyperref}
\newsavebox{\tempbox}
\usepackage{epstopdf}
\usepackage{booktabs}
\newcommand{\bigcell}[2]{\begin{tabular}{@{}#1@{}}#2\end{tabular}}

\usepackage{geometry}
\usepackage{changepage}
\usepackage{numprint}  
\usepackage{longtable} 
\usepackage{array}
\usepackage{color}
\makeatletter
\usepackage{epstopdf}

\makeatother

\usepackage{atlaspackage}

\usepackage{atlasphysics}

\usepackage{ScalarLQPaper-defs}

\hypersetup{pdftitle={ATLAS draft},pdfauthor={The ATLAS Collaboration}}

\input{ScalarLQPaper-metadata}

\begin{document}

\tableofcontents

\newpage

\input{LQIntro.tex}
\input{Detector.tex}

\input{DataBG.tex}

\input{FirstSecondGen.tex}

\input{Sbottom.tex}
\input{Stop.tex}

\input{Conclusions.tex}

\section*{Acknowledgements}
\input{Acknowledgements.tex}

\bibliographystyle{atlasBibStyleWoTitle}
\input{ScalarLQPaper.bbl}

\newpage \input{atlas_authlist}

\end{document}

%% file: ScalarLQPaper-metadata.tex
\usepackage{ulem}
\usepackage{subfig}
\usepackage{lscape}
\usepackage{verbatim}
\usepackage{float}
\usepackage{graphicx}
\usepackage{url}
\usepackage{caption}
\usepackage{hyperref}
\usepackage{epstopdf}
\usepackage{booktabs}
\usepackage{cite}
\usepackage{geometry}
\usepackage{changepage}
\usepackage{numprint}  
\usepackage{longtable} 
\usepackage{array}
\usepackage{color}
\makeatletter

\usepackage{epstopdf}

\makeatother

\AtlasTitle{Searches for scalar leptoquarks in $pp$ collisions at \\$\boldmath{\sqrt{s}}$ = 8\,TeV with the ATLAS detector}

\author{The ATLAS Collaboration}

\PreprintIdNumber{CERN-PH-EP-2015-179}
\arXivId{15XX.YYYY}

\AtlasAbstract{Searches for pair-produced scalar leptoquarks are performed using 20~fb$^{-1}$ of proton--proton collision data provided by the LHC and recorded by the {ATLAS} detector at $\sqrt{s}=8$~TeV. Events with two electrons (muons) and two or more jets in the final state are used to search for first (second)-generation leptoquarks. The results from two previously published ATLAS analyses are interpreted in terms of third-generation leptoquarks decaying to $b\nu_{\tau}\bar{b}\bar{\nu_{\tau}}$ and $t\nu_{\tau}\bar{t}\bar{\nu_{\tau}}$ final states. No statistically significant excess above the Standard Model expectation is observed in any channel and scalar leptoquarks are excluded at 95\% CL with masses up to $\mlqone<$~1050~GeV for first-generation leptoquarks, $\mlqtwo<$~1000~GeV for second-generation leptoquarks, $\mlqthree<$~625~GeV for third-generation leptoquarks in the $b\nu_{\tau}\bar{b}\bar{\nu_{\tau}}$~channel, and 200~$< \mlqthree < $~640~GeV in the $t\nu_{\tau}\bar{t}\bar{\nu_{\tau}}$~channel.}

%% file: LQIntro.tex
\section{Introduction}
Leptoquarks (LQ) are predicted by many extensions of the Standard Model (SM) \cite{BSM,techni2,techni3,string,comp,pati_salam_colour,georgi_glashow_unification}~and may provide an explanation for the many observed similarities between the quark and lepton sectors of the SM.  LQs are colour-triplet bosons with fractional electric charge.  They carry non-zero values of both baryon and lepton number \cite{Schrempp:1984nj}. They can be either scalar or vector bosons and are expected to decay directly to lepton--quark pairs (where the lepton can be either charged or neutral). 

The coupling strength between scalar LQs and the lepton-quark pairs depends on a single Yukawa coupling, termed $\lambda_{\mathrm{LQ} \rightarrow \ell q}$. The additional magnetic moment and electric quadrupole moment interactions of vector LQs are governed by two coupling constants \cite{Hewett:1997ce}. The coupling constants for both the scalar and vector LQs, and the branching fraction of the LQ decay into a quark and a charged lepton, $\beta$, are model dependent.  The production cross-section and couplings of vector LQs are enhanced relative to the contribution of scalar LQs, although the acceptance is expected to be similar in both cases.  This analysis considers the simpler scenario of scalar LQ pair-production, for which the form of the interaction is known and which provides more conservative limits on LQ pair-production than for vector LQ pair-production. 

In proton--proton collisions, LQs can be produced singly and in pairs. The production of single LQs, which happens at hadron colliders in association with a lepton, depends directly on the unknown Yukawa coupling $\lambda_{\mathrm{LQ} \rightarrow \ell q}$.  However, LQ pair-production is not sensitive to the value of the coupling.  In \textit{pp} collisions with a centre-of-mass energy $\sqrt{s}$ = 8\,\TeV, the dominant pair-production mechanism for LQ masses below $\sim$1~TeV is gluon fusion, while the \textit{qq}-annihilation production process becomes increasingly important with increasing LQ mass. 

The minimal Buchm\"uller--R\"uckl--Wyler model (mBRW) \cite{Buchmuller:1986zs} is used as a benchmark model for scalar LQ production. It postulates additional constraints on the LQ properties, namely that the couplings have to be purely chiral, and makes the assumption that LQs are grouped into three families (first, second and third-generation) that couple only to leptons and quarks within the same generation. The latter requirement excludes the possibility of flavour-changing neutral currents (FCNC) \cite{Mitsou:2004hm}, which have not been observed to date.

Previous searches for pair-produced LQs have been performed by the ATLAS Collaboration with 1.03~fb$^{-1}$ of data collected at $\sqrt{s}$~=~7~TeV, excluding at 95\% confidence level (CL) the existence of scalar LQs with masses below 660~(607)~GeV for first-generation LQs at $\beta=1$~(0.5) \cite{Aad:2011ch} and  685~(594)~GeV for second-generation LQs at $\beta=1$~(0.5) \cite{ATLAS:2012aq}.  The CMS Collaboration excluded at 95\% CL the existence of scalar LQs with masses below 830~(640)~GeV for first-generation LQs at $\beta=1$~(0.5) and 840~(650)~GeV for second-generation LQs at $\beta=1$~(0.5) with 5.0~fb$^{-1}$ of data collected at $\sqrt{s}$~=~7~\TeV \cite{Chatrchyan:2012vza}.

Pair-produced third-generation scalar LQs decaying to \bbMET~have been excluded by the CMS Collaboration for masses below 700\,\GeV\,at $\beta$\ = 0, and for masses below 560\,\GeV\, over the full $\beta$ range using 19.7~fb$^{-1}$ of data collected at $\sqrt{s}$~=~8~TeV \cite{Khachatryan:2015bsa}.  Third-generation scalar LQs have been excluded in the $b\tau^{+}\bar{b}\tau^{-}$~channel at $\beta = 1$ for masses up to 740\,\GeV\ by the CMS Collaboration using 19.7~fb$^{-1}$ of data collected at $\sqrt{s}$~=~8~TeV \cite{CMS_LQ3_bbtautau}, and by the ATLAS Collaboration at $\beta = 1$ for masses up to 534 \GeV\ using 4.7~fb$^{-1}$ of data collected at $\sqrt{s}$~=~7~TeV \cite{ATLAS_LQ3_bbtautau}.  The CMS Collaboration also excluded third-generation scalar LQs in the $t\tau^{-}\bar{t}\tau^{+}$~channel at $\beta = 1$ for masses up to 685\,\GeV\, using 19.7~fb$^{-1}$ of data collected at $\sqrt{s}$~=~8~TeV \cite{Khachatryan:2015bsa}.  

In this paper, searches for pair-produced first- and second-generation scalar LQs (\LQone\,and \LQtwo, respectively) are performed by selecting events with two electrons or muons plus two jets in the final state (denoted by \eejj~and \mumujj, respectively). In addition, limits are placed on pair-produced third-generation scalar LQs (\LQthree) by reinterpreting ATLAS searches for supersymmetry (SUSY) in two different channels \cite{ATLAS_sbottom, ATLAS_stop}.  LQ production and decay mechanisms can be similar to those of stop quarks (\stop) and sbottom quarks~(\sbottom).  For example, $\stop \stop \rightarrow tt \tilde{\chi}^{0} \tilde{\chi}^{0}$ gives the same event topology as $\LQLQ\rightarrow$\ttMET\ in the limit where the neutralino ($\tilde{\chi}^{0}$) is massless.  Two ATLAS analyses optimised for these SUSY processes are therefore used to set limits on the equivalent LQ decay processes: $\LQLQ\rightarrow$\bbMET\ and $\LQLQ\rightarrow$\ttMET.

The results for each \LQthree\ channel cannot be combined since the parent LQs have different electric charges in the two cases ($-\frac{1}{3}e$~for the $\LQLQ\rightarrow$\bbMET~channel and $\frac{2}{3}e$~for the $\LQLQ\rightarrow$\ttMET\ channel, where $e$~is the elementary electric charge). The branching fractions of \LQthree~decays to $b\nutau$~and $t\nutau$~are assumed to be equal to 100\% in each case.  Although complementary decays of a charge $-\frac{1}{3}e$~($\frac{2}{3}e$) LQ into a $t\tau^{-}\bar{t}\tau^{+}$~($b\tau^{+}\bar{b}\tau^{-}$) final state are also allowed, kinematic suppression factors which favour LQ~decays to $b$-quarks over $t$-quarks and the relative strengths of the Yukawa couplings would have to be considered.  Since these suppression factors are model dependent, limits are not provided as a function of $\beta$ for the \LQthree\ channels.

%% file: Detector.tex
\section{The ATLAS detector}

The ATLAS experiment \cite{det2} is a multi-purpose  detector with a forward-backward symmetric cylindrical geometry and nearly 4$\pi$ coverage in solid  angle.  The three major sub-components of ATLAS are the tracking detector, the calorimeter and the muon spectrometer.  Charged-particle tracks and vertices are reconstructed by the inner detector (ID) tracking system, comprising silicon pixel and microstrip detectors covering the pseudorapidity\footnote{ATLAS uses a right-handed coordinate system with its origin at the nominal interaction point (IP) in the centre of the detector and the $z$-axis along the beam pipe. The $x$-axis points from the IP to the centre of the LHC ring, and the $y$-axis points upward. Cylindrical coordinates $(r,\phi)$ are used in the transverse plane, $\phi$ being the azimuthal angle around the beam pipe. The pseudorapidity is defined in terms of the polar angle $\theta$ as $\eta=-\ln\tan(\theta/2)$.} range $|\eta|$~$<$~2.5, and a straw tube tracker that covers $|\eta|$~$<$~2.0.  The ID is immersed in a homogeneous 2\,T magnetic field provided by a solenoid.  Electron, photon, jet and tau energies are measured with sampling calorimeters.  The ATLAS calorimeter system covers a pseudorapidity range of $|\eta|$~$<$~4.9.  Within the region $|\eta|$~$<$~3.2, electromagnetic calorimetry is provided by barrel and endcap high-granularity lead/liquid argon (LAr) calorimeters, with an additional thin LAr presampler covering $|\eta|$~$<$~1.8, to correct for energy loss in material upstream of the calorimeters. Hadronic calorimetry is provided by a steel/scintillator-tile calorimeter, segmented into three barrel structures within $|\eta|$~$<$~1.7, and two copper/LAr hadronic endcap calorimeters. The forward region (3.1~$<$~$|\eta|$~$<$~4.9) is instrumented by a LAr calorimeter with copper (electromagnetic) and tungsten (hadronic) absorbers.  Surrounding the calorimeters is a muon spectrometer (MS) with air-core toroids, a system of precision tracking chambers providing coverage over $|\eta|$~$<$~2.7, and detectors with triggering capabilities over $|\eta|$~$<$~2.4 to provide precise muon identification and momentum measurements.

%% file: DataBG.tex
\section{Data and Monte Carlo samples}
\label{sec:DataMC}

The results presented here are based on proton--proton collision data at a centre-of-mass energy of \sqrts~=~8\,\TeV, collected by the ATLAS detector at the LHC during 2012. Data samples corresponding to an integrated luminosity of 20.3\,\ifb~are used by all channels except for the $\LQLQ\rightarrow\bbMET$~analysis which uses 20.1\,\ifb~because of requirements made by the trigger used in the analysis.

Simulated signal events of pair-produced scalar LQs decaying to $e^{+}e^{-}q\bar{q}$, $\mu^{+}\mu^{-}q\bar{q}$, \ttMET, and \bbMET~final states are  produced  using the \pythia 8.160 \cite{Sjostrand:2007gs} event generator with  \texttt{CTEQ6L1}~\cite{CTEQ}  parton distribution  functions (PDFs).  The coupling $\lambda_{\mathrm{LQ} \rightarrow \ell q}$ which determines the LQ lifetime and width \cite{LQlq_coupling} is set  to $\sqrt{0.01\times4\pi\alpha}$, where $\alpha$ is the fine-structure constant. This value gives the LQ a full width of less than 100\,\MeV, which is much smaller than the detector resolution. For LQ masses in the ranges considered here (200~\GeV\,$\le$\,$m_{\mathrm{LQ}}$\,$\le$~1300~\GeV, in steps of 50~GeV), the value of the coupling used is such that the LQs can be considered to decay promptly. The production cross-section of pair-produced LQs is assumed to be independent of the coupling strength. The signal process is normalised to the expected next-to-leading-order (NLO) cross-sections for scalar LQ pair-production \cite{Kramer:2004df}.  The signal production cross-section is 23.5~fb for a LQ mass of 600 \GeV, and 0.40~fb for a 1~\TeV~LQ and is the same for each generation.

\subsection{Monte Carlo for background predictions}
The Monte Carlo (MC) samples used to estimate the contributions from SM backgrounds to the \LQone\ and \LQtwo\ searches are discussed here.  Details about the MC models used for estimating backgrounds in the \LQthree\ searches are available in Refs. \cite{ATLAS_sbottom}\ (for the \bbMET\ channel) and \cite{ATLAS_stop}\ (for the \ttMET\ channel).

The MC samples used to model the $Z/\gamma^*$+jets background with a dilepton invariant mass ($\mll$) less than 120~GeV and high-mass Drell--Yan backgrounds ($\mll\geq$~120~GeV) are generated with \texttt{SHERPA 1.4.1} \cite{Gleisberg:2008ta}.  The high-mass Drell--Yan samples are generated assuming massive $c$- and $b$-quarks instead of the conventional massless treatment.

Samples of $t\bar{t}$\ events are produced with \texttt{POWHEG} box \cite{Nason:2004rx,Frixione:2007vw} interfaced with \texttt{PYTHIA 6}.
MC samples representing the $WW$, $WZ$, and $ZZ$\ diboson decays are generated with \texttt{HERWIG 6.52} \cite{Corcella:2000bw} and use the \texttt{AUET2} \cite{AUET2} values for the tunable parameters (the `\texttt{AUET2} MC tune'). Samples of single-top-quark events in the $Wt$ and $s$-channel are generated with \texttt{MC@NLO 4.01} \cite{Frixione:2005vw,Frixione:2008yi} and the \texttt{AUET2} MC tune, while the $t$-channel samples are generated with \texttt{AcerMC 3.8} \cite{Kersevan:2004yg} interfaced with \texttt{PYTHIA 8} and use the \texttt{AUET2B} \cite{ATLAS:2011krm} MC tune. The hadronisation and parton showering of the samples produced with \texttt{MC@NLO} are done using \texttt{HERWIG 6.52} coupled to \texttt{JIMMY 4.31} \cite{Butterworth:1996zw}. The $W$+jets samples are produced with \texttt{ALPGEN 2.14} interfaced with \texttt{JIMMY 4.31}, also with the \texttt{AUET2} MC tune applied. 
The choice of PDFs used to produce the MC simulated samples is generator dependent: \texttt{AcerMC}, \texttt{PYTHIA}, \texttt{HERWIG} and \texttt{ALPGEN} use \texttt{CTEQ6L1}, while \texttt{MC@NLO} uses \texttt{CT10} \cite{Lai:2010vv}. For all samples, the detector response is modelled \cite{ATLAS_Simulation} using \texttt{GEANT4} \cite{GEANT4}, except for the Drell--Yan background samples, which use a fast detector simulation where the calorimeter response is parameterised. The differences between fast and full simulation in terms of kinematic spectra and modelling of relevant objects are evaluated to be negligible.

The cross-sections of background processes used in the analysis are taken from theoretical predictions.  Single-top production cross-sections in the $s$-channel \cite{Kidonakis:2010tc}, $t$-channel \cite{Kidonakis:2011wy}, and in associated production with a $W$ boson \cite{Kidonakis:2010ux}, are calculated to NLO+NNLL accuracy. $W$+jets and $Z\rightarrow\tau\tau$ cross-sections with NNLO accuracy are used \cite{Catani:2009sm}. The cross-sections for $WW$, $WZ$, and $ZZ$\ processes are calculated at NLO \cite{Campbell:2011bn,Campbell:1999ah}.  The theoretical cross-section for $WW$\ production is scaled by a factor 1.2 and the uncertainty is increased by an extra 20\%, in order to take into account the ATLAS \cite{ATLAS:2012mec} and CMS measurements \cite{Chatrchyan:2012bd}, which showed an excess in data at the level of 20\%\ (see Refs.~\cite{Gehrmann:2014fva,Jaiswal:2014yba} for more discussion about possible causes of the excess). 

For the $Z/\gamma^*$+jets and $t\bar{t}$ backgrounds, LO and NLO cross-sections, respectively are used. These backgrounds are constrained using two control regions (CRs), as described in Section~\ref{sec::CtrlRegion}.

%% file: FirstSecondGen.tex
\section{Searches for first- and second-generation LQs}

The first- and second-generation analyses exploit similarities in the final states and use common search strategies to select dilepton plus dijet final states.  Control regions are used to constrain estimates of the dominant backgrounds to the data.  A set of discriminating variables is used to define signal regions (SRs) that are used for a counting analysis. 

\subsection{Trigger and data collection}
\label{subsec:trigger}

Selected data events are required to have all relevant components of the ATLAS detector in good working condition. For the \LQone~(\eejj) analysis, the trigger requires at least two electromagnetic calorimeter clusters, defined as energy deposits in the cells of the electromagnetic calorimeter. The leading cluster is required to have transverse momentum $\pt$~>~$35$~GeV and the sub-leading one $\pt$~>~$25$~GeV.   This trigger selects electrons without imposing any requirement on the isolation and this allows a data-driven estimate of the background contribution from jets in the final state that pass the electron selection, as described in detail in Ref.~\cite{ZprimeNote}. The trigger is 98\% efficient with respect to the offline selection, which requires $\pt$ above 40~(30)~GeV for the leading (sub-leading) electron.

For the \LQtwo~(\mumujj) analysis, events are selected from data using a trigger which requires the presence of at least one muon candidate in the event with \pt\ above 36~\GeV. This trigger is fully efficient relative to the offline selection for muons with $\pt$ above 40~GeV \cite{Aad:2014sca}.  

\subsection{Object selection}
\label{sec:ObjSel}

Electrons are selected and identified by imposing requirements on the shape of the cluster of energy deposits in the calorimeter, as well as on the quality of the track, and on the track-to-cluster matching.  The identification efficiency is on average 85\% \cite{el_id_2014}.  Electron candidates must have transverse energy $\ET>30$\,\GeV~and $|\eta|<2.47$.  Electron candidates associated with clusters in the transition region between the barrel and endcap calorimeters ($1.37$~<~$|\eta|$~<~$1.52$) are excluded.  All electrons are required to be reconstructed with cluster-based or combined cluster- and track-based algorithms and to satisfy calorimeter quality criteria. Requirements are made on the transverse ($|d_0|$) and longitudinal ($|z_0|$) impact parameters of the electron relative to the primary vertex and must satisfy $|d_0| < 1$~mm and $|z_0| < 5$~mm.
In addition, electrons are required to be isolated by imposing requirements on the $\etcone$ measured in the calorimeter within a cone of size $\Delta R=\sqrt{(\Delta\eta)^2 + (\Delta\phi)^2} = 0.2$ around the electron cluster excluding the electron cluster energy, and corrected to account for leakage (i.e. energy deposited by the electron outside of the cluster) and the average number of proton--proton interactions per bunch-crossing. The isolation requirements are optimised for high-\pt~electrons following the strategy in Ref.~\cite{ZprimeNote}.  The leading electron is required to have $\etcone<0.007\times\ET+5$~\GeV, and the sub-leading electron is required to have $\etcone<0.022\times\ET+6$~\GeV.

Muon tracks are reconstructed independently in the ID and the MS. Tracks are required to have a minimum number of hits in each system, and must be compatible in terms of geometrical and momentum matching. In particular, in order to prevent mis-measurements at high $\pt$, muons are required to have hits in all three MS stations, as described in Ref.~\cite{ZprimeNote}.  In order to increase the muon identification efficiency, when one muon in the event satisfies the three-stations requirement, the criteria for the second muon in the event are relaxed to require hits in only two MS stations.
Information from both the ID and MS is used in a combined fit to refine the measurement of the momentum of each muon \cite{Aad:2014rra}.  Muon candidates are required to have $\pt>40$~GeV, $|\eta|$~<~$2.4$, $|d_0|<0.2$~mm and $|z_0|<1.0$~mm.  Muons must also pass a relative-isolation requirement $\ptcone/\pt<0.2$, where $\ptcone$ is the sum of the transverse momenta of all the tracks with $\pt$ above 1~GeV (except for the muon track) within a cone of $\Delta R$~<~$0.2$ around the muon track, and $p_{\mathrm{T}}$ is the transverse momentum of the muon. 

Jets are reconstructed from clusters of energy deposits detected in the calorimeter using the anti-$k_{t}$ algorithm \cite{AntiKt} with a radius parameter $R=0.4$ \cite{Lampl:1099735}. They are calibrated using energy- and $\eta$-dependent correction factors derived from simulation and with residual corrections from in-situ measurements.
The jets used in the analysis must satisfy $\pt$~>~$30$~GeV and $|\eta|$~<~$2.8$. Jets reconstructed within a cone of $\Delta R=0.4$ around a selected electron or muon are removed. Additional jet quality criteria are also applied to remove fake jets caused by detector effects. A detailed description of the jet energy scale measurement and its systematic uncertainties is given in Ref.~\cite{Aad:2014bia}. 

\subsection{Event pre-selection}
\label{sec:EvtSel}

Multiple $pp$ interactions during bunch-crossings (pile-up) can give rise to multiple reconstructed vertices in events. The primary vertex of the event, from which the leptons are required to originate, is defined as the one with the largest sum of squared transverse momenta of its associated tracks. Events are selected if they contain a primary vertex with at least three associated tracks satisfying $p_{\mathrm{T,track}} >$ 0.4 GeV.

MC events are corrected to better describe the data by applying a per-event weight to match the distribution of the average number of primary vertices observed in data. A weighting factor is also applied in order to improve the modelling of the vertex position in $z$. Scale factors are applied to account for differences in lepton identification and selection efficiency between data and MC simulation. The scale factors depend on the lepton kinematics and are described in detail in Ref.~\cite{Aad:2014rra} for muons, and in Ref.~\cite{Aad:2014nim} for electrons. The energy and momentum of the selected physics objects are corrected to account for the resolution and scale measured in data, as described in Ref.~\cite{Aad:2014rra} for muons, in Ref.~\cite{Aad:2014nim} for electrons and in Ref.~\cite{Aad:2014bia} for jets.

Events are selected in the \eejj~channel if they contain exactly two electrons with $\pt$~>~40~(30)~GeV for the leading (sub-leading) electron and at least two jets with $\pt$~>~30~GeV.  For the \mumujj~channel, events are selected if they contain exactly two muons with \pt\,$>$\,40\,\GeV\,and opposite-sign charge, and at least two jets with \pt\,$>$\,30\,\GeV.  No requirements are placed on the charges of the electron candidates due to inefficiencies in determining the charge of high-\pt~tracks associated with electrons.  These sets of requirements form the basic event `pre-selection' for the analyses, which is used to build the control and signal regions discussed in the following sections.  

\subsection{Signal regions}
\label{sec:SignalRegions}

After applying the event pre-selection requirements, a set of signal regions is defined using additional kinematic variables in order to discriminate LQ signals from SM background processes and to enhance the signal-to-background ratio. The variables used are: 
\begin{itemize}
\item[--] \mll: The dilepton invariant mass.
\item[--] $S_{\mathrm{T}}$: The scalar sum of the transverse momenta of the two leading leptons and the two leading jets.
\item[--] $\mlqmin$:  The lowest reconstructed LQ mass in the event.  The reconstructed masses of the two LQ candidates in the event (\mlqmin~and \mlqmax) are defined as the invariant masses of the two lepton--jet pairs with the smallest difference (and $\mlqmin < \mlqmax$).
\end{itemize}

Signal regions are determined by optimising the statistical significance as defined in Ref.~\cite{Cowan:2010js}.  The optimisation procedure is performed in a three-dimensional space constructed by \mll, \st~and \mlqmin~~for each of the signal mass points. Several adjacent mass points may be grouped into a single SR.  The signal acceptance of the selection requirements is estimated to be $\approx50$\% in the \mumujj~channel and between 65\% and 80\% in the \eejj~channel (assuming $\beta$\ = 1.0).  The difference is due to tighter quality selection requirements in the \mumujj~channel used to prevent muon mis-measurements in MS regions with poor alignment or missing chambers. The optimised signal regions are presented in Table~\ref{tab:LQ12_SRDefs} together with the mass of the corresponding LQ hypothesis. Each LQ mass hypothesis is tested in only one signal region, where limits on $\sigma\times\beta$\ are extracted.

\begin{table}[htbp]
\footnotesize
  \begin{center}
    \begin{tabular}{ c | c | c c c }
      \hline 
      &
      \bigcell{c}{LQ masses \\ $[$GeV$]$} &
      \bigcell{c}{$\mll$ \\ $[$GeV$]$} & 
      \bigcell{c}{$\st$ \\$[$GeV$]$} &
      \bigcell{c}{$\mlqmin$ \\ $[$GeV$]$} \\ 
      \hline 
      SR1  &    300       &    130  &    460    &    210  \\ 
      SR2  &    350       &    160  &    550    &    250  \\ 
      SR3  &    400       &    160  &	 590    &    280  \\ 
      SR4  &    450       &    160  &	 670    &    370  \\ 
      SR5  &    500--550  &    180  &    760    &    410  \\ 
      SR6  &    600--650  &    180  &    850    &    490  \\ 
      SR7  &    700--750  &    180  &    950    &    580  \\ 
      SR8  &    800--1300 &    180  &    1190   &    610  \\ \hline 
    \end{tabular}
  \end{center}
  \captionof {table}{The minimum values of \mll, \st, and \mlqmin~used to define each of the signal regions targeting different LQ masses in the \eejj\ and \mumujj\ channels. Each signal region is valid for one or more mass hypotheses, as shown in the second column.}
  \label{tab:LQ12_SRDefs}
\end{table}

\subsection{Background estimation}
\label{sec::CtrlRegion}

The main SM background processes to the \LQone\ and \LQtwo\ searches are the production of $Z/\gamma^*$+jets events, $t\bar{t}$ events where both top quarks decay leptonically, and diboson events. Additional small contributions are expected from $Z\rightarrow\tau\tau$ and single-top processes. Multi-jets, $W$+jets, $t\bar{t}$ (where one or more top quarks decays hadronically), and single-top events with mis-identified or non-prompt leptons arising from hadron decays or photon conversions can also contribute. These fake lepton backgrounds are estimated separately in the \eejj~and \mumujj~channels using the same data-driven techniques as described in Ref.~\cite{ZprimeNote}~and are found to be negligible for the \mumujj~channel. Normalisation factors, derived using background-enriched control regions, are applied to the MC predictions for $Z/\gamma^*$+jets and $t\bar{t}$~backgrounds to predict as accurately as possible the background in the signal regions.  These control regions are constructed to be mutually exclusive to the signal region and the assumption is made that normalisation factors and their associated uncertainties in the signal region are the same as in the background-enriched control regions.

\subsubsection{Control regions for $Z/\gamma^*$+jets and $t\bar{t}$~backgrounds}
\label{sec::CtrlRegion::CRs}

Two control regions with negligible signal contributions are defined to validate the modelling accuracy of the MC simulated background events and to derive normalisation scale factors. The $Z/\gamma^*$+jets control region is defined by the pre-selection requirements with an additional requirement of 60 < $\mee$\ < 120 \GeV\ in the \eejj~channel and 70 < $\mmumu$\ < 110 \GeV\ in the \mumujj~channel. These control regions define a pure sample of $Z/\gamma^*$+jets events. The \ttbar\ control region is defined in both channels by applying the pre-selection requirements, but demanding exactly one muon and one electron (both with \pt~above 40 \GeV) in the offline selection instead of two same-flavour leptons.  In the case of the $t\bar{t}$ control region for the \eejj\ channel, the trigger requirement is modified by requiring a single isolated electron with $\pt$\ above $24$ \GeV, which is fully efficient relative to the offline selection for electrons with $\pt$ above 30 \GeV.  In both cases, the same selection criteria are applied to data and MC events.

Normalisation factors are applied to the MC predictions for the $Z/\gamma^*$+jets and $t\bar{t}$ background processes. They are obtained by performing a combined maximum likelihood fit to the observed yields in the control regions and signal region under consideration. Systematic uncertainties on the predicted MC yields related to the uncertainty on the cross-sections are taken into account by the fit through the use of dedicated nuisance parameters.  The fit procedure is performed using the {\sc{HistFitter}} package \cite{Baak:2014wma}, which is a tool based on the {\sc{RooStats}} framework \cite{Moneta:2010pm}. The normalisation scale factor obtained from a background-only fit for the $Z/\gamma^*$+jets background in the \eejj~(\mumujj) channel is $1.1\pm0.2$ ($0.97\pm0.15$), while the normalisation scale factor for $t\bar{t}$~is $1.10\pm0.05$ ($1.01\pm0.05$). The fitted background scale factors have little sensitivity to the inclusion of signal regions and the eventual presence of a signal.

\subsubsection{Kinematic distributions}
\begin{figure}[htbp]
  \centering
  \subfloat[\eejj channel]{
    \includegraphics[width=0.5\textwidth]{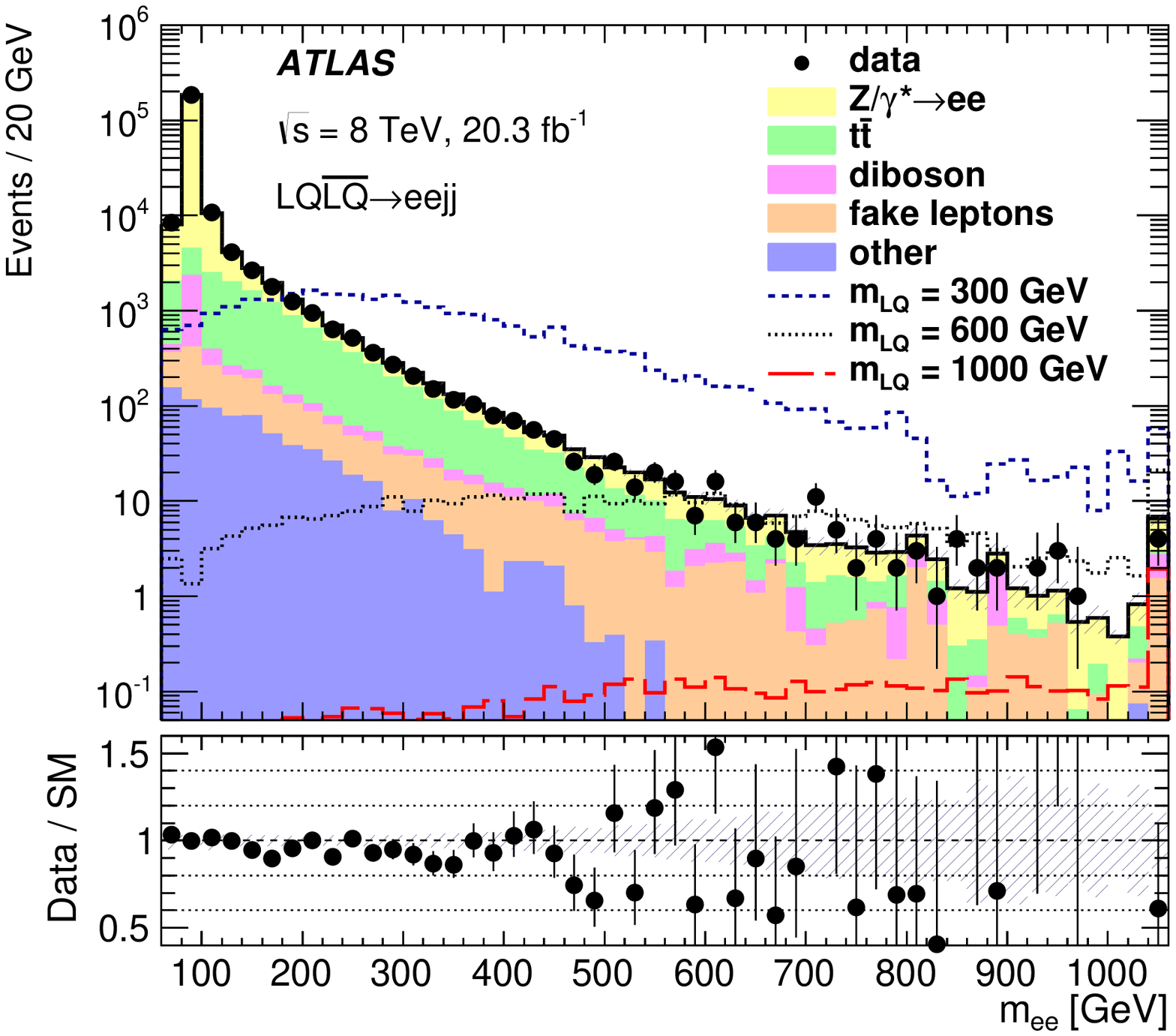}
    \label{fig:lq1_mll}
  }
  \subfloat[\mumujj channel]{
    \includegraphics[width=0.5\textwidth]{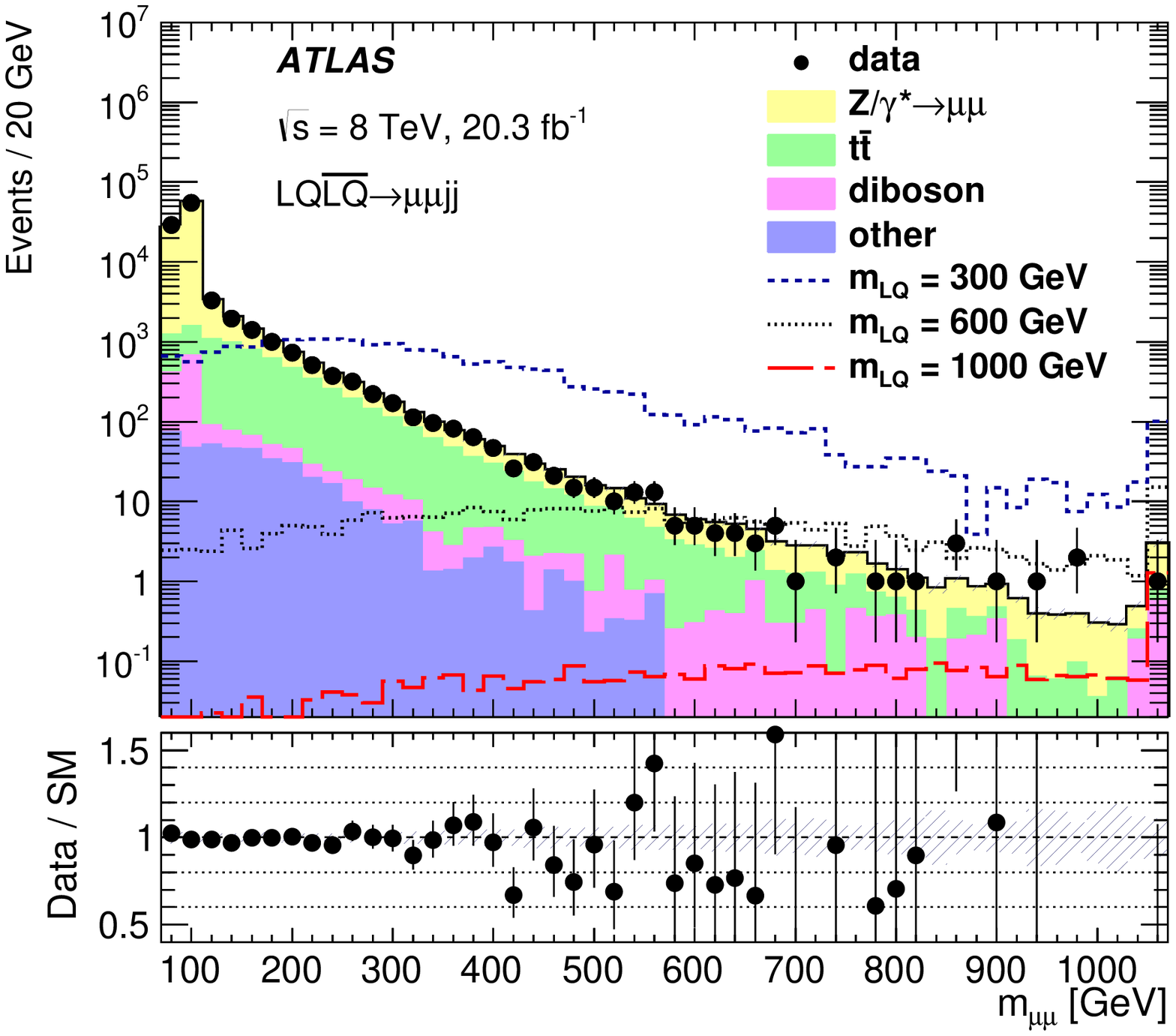}
    \label{fig:lq2_mll}
  }
  \caption{Distributions of the dilepton invariant mass (\mll) in the \eejj~(left) and \mumujj~(right) channels after applying the pre-selection cuts.  The signal model assumes $\beta$ = 1.0.  The last bin includes overflows. The ratio of the number of data events to the number of background events (and its statistical uncertainty) is also shown.  The hashed bands represent all sources of statistical and systematic uncertainty on the background prediction.}
  \label{fig:mll}
\end{figure}

\begin{figure}[htbp]
  \centering
  \subfloat[\eejj channel]{
    \includegraphics[width=0.5\textwidth]{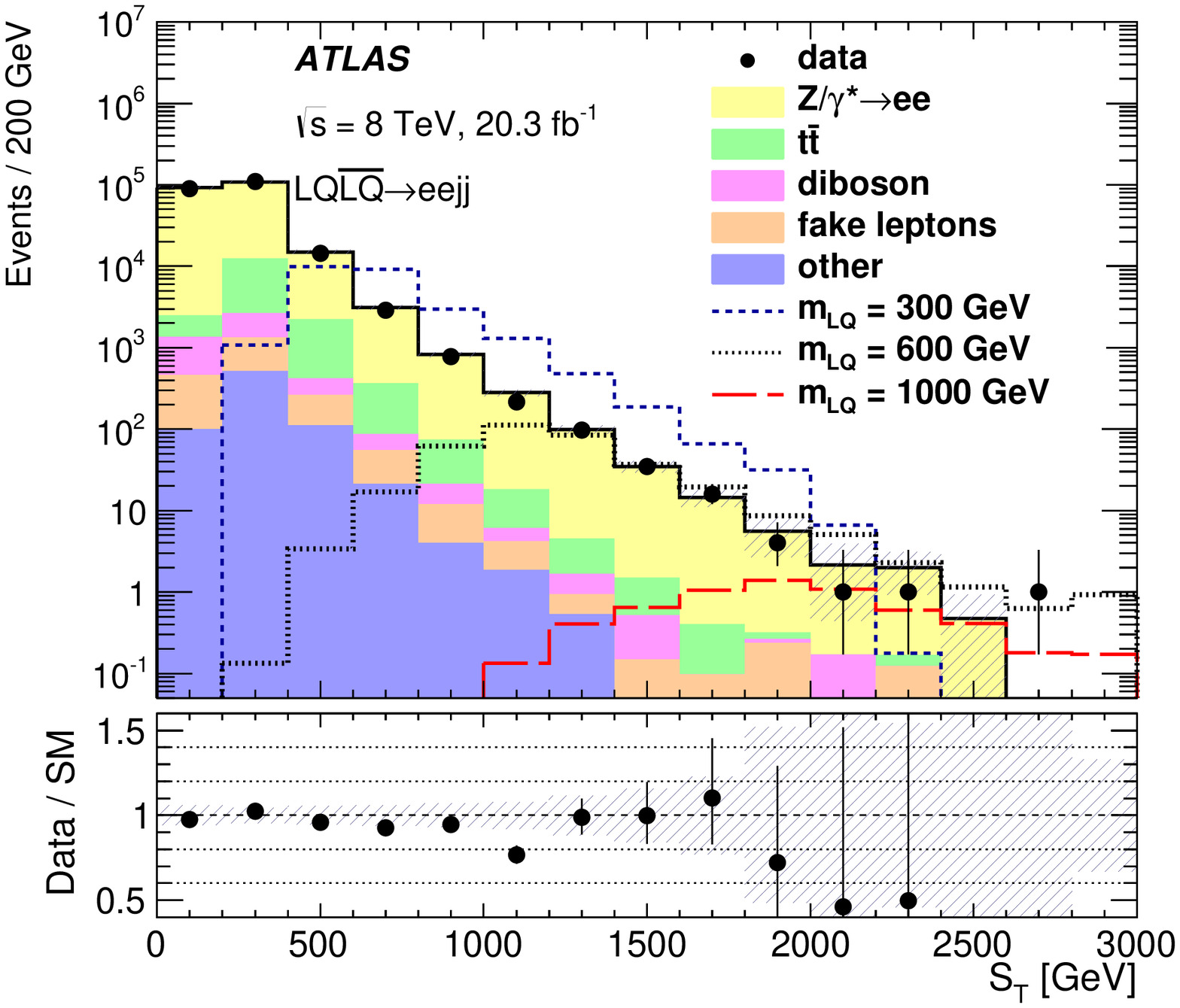}
    \label{fig:lq1_st}
  }
  \subfloat[\mumujj channel]{
    \includegraphics[width=0.5\textwidth]{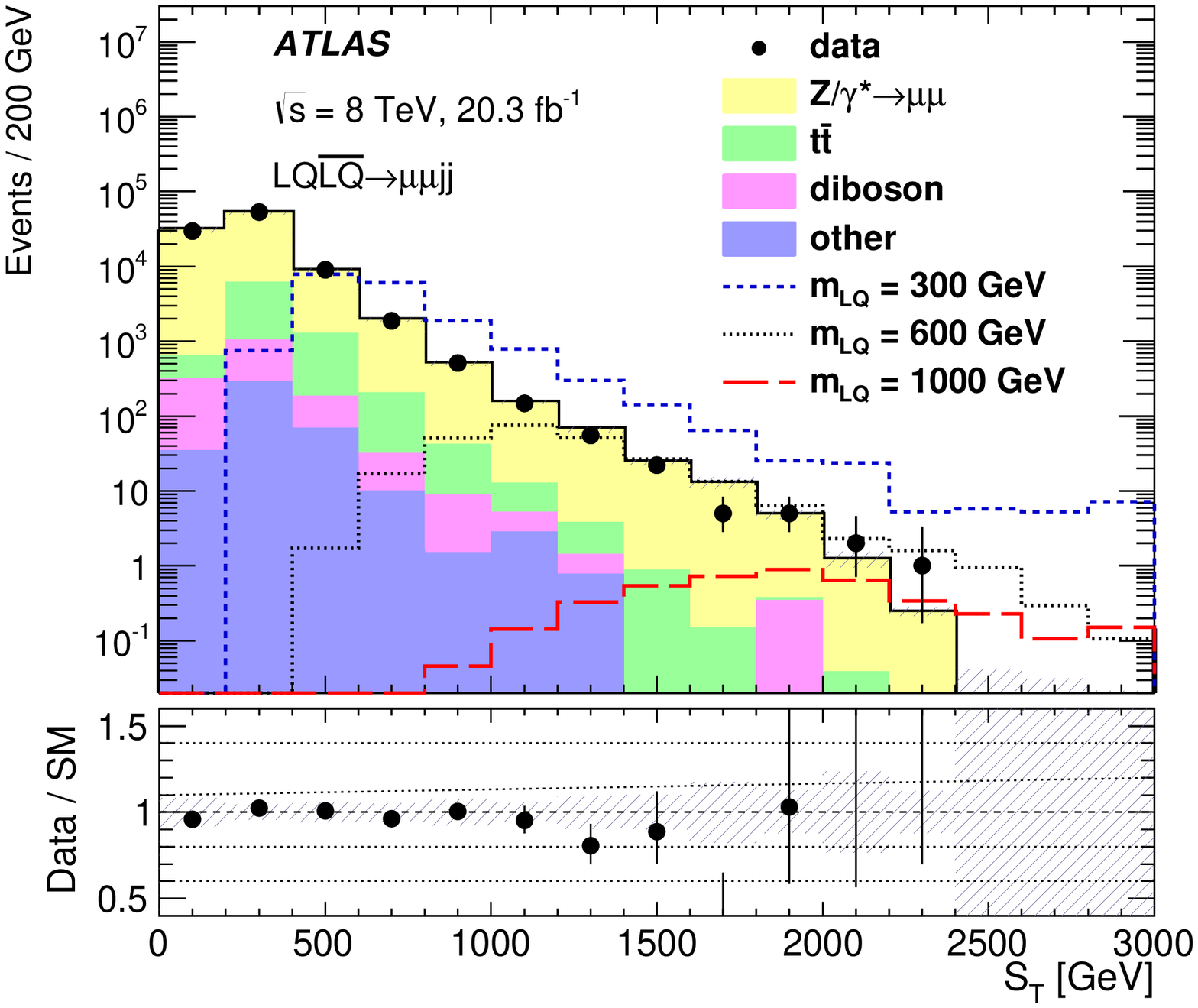}
    \label{fig:lq2_st}
  }
  \caption{Distributions of the total scalar energy (\st) in the \eejj~(left) and \mumujj~(right) channels after applying the pre-selection cuts.  The signal model assumes $\beta$\ = 1.0.  The last bin includes overflows. The ratio of the number of data events to the number of background events (and its statistical uncertainty) is also shown.  The hashed bands represent all sources of statistical and systematic uncertainty on the background prediction.}
  \label{fig:st}
\end{figure}

\begin{figure}[htbp]
  \centering
  \subfloat[\eejj channel]{
    \includegraphics[width=0.5\textwidth]{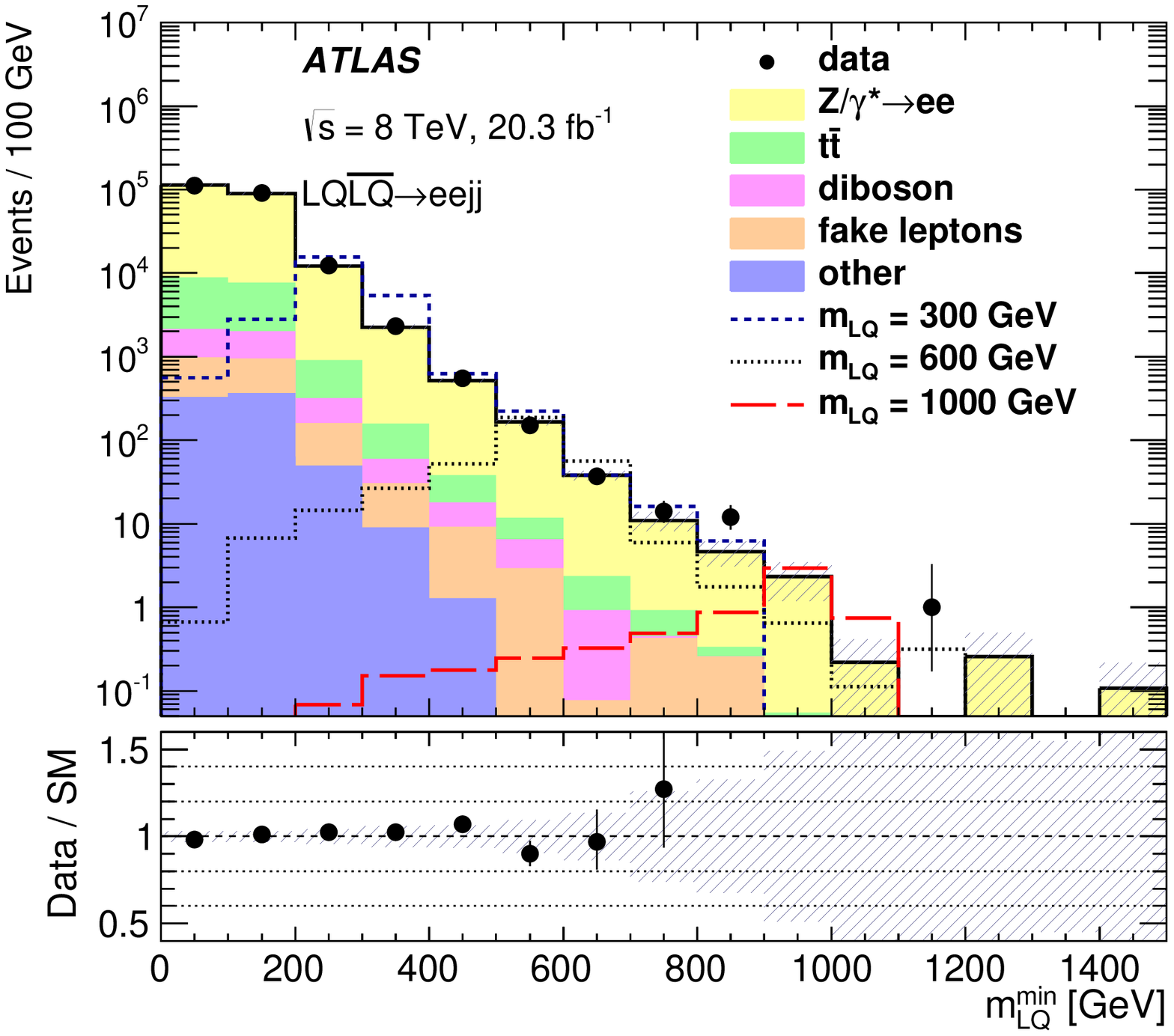}
    \label{fig:lq1_mlq}
  }
  \subfloat[\mumujj channel]{
    \includegraphics[width=0.5\textwidth]{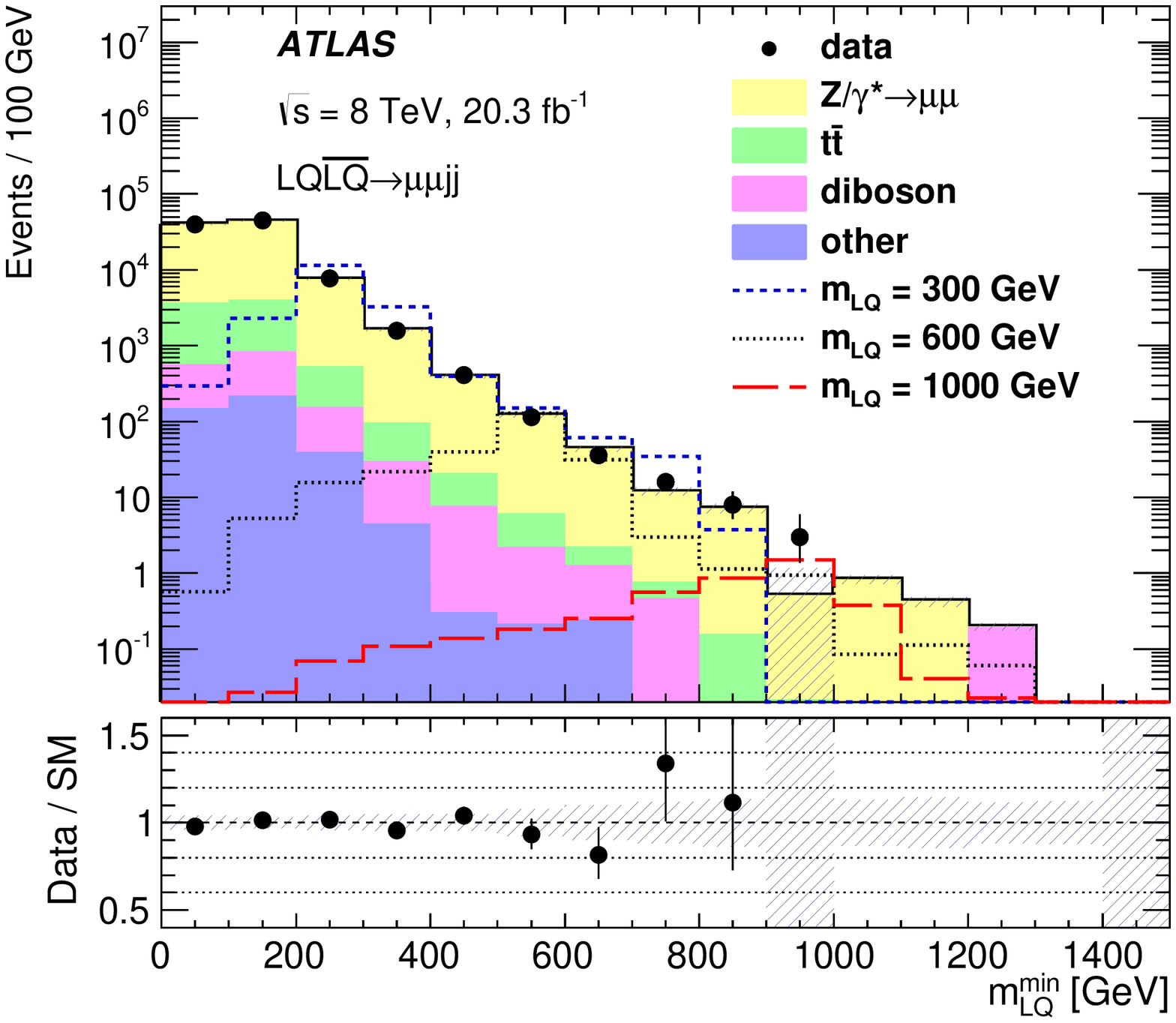}
    \label{fig:lq2_mlq}
  }
  \caption{Distributions of the lowest reconstructed LQ mass (\mlqmin) in the \eejj~(left) and \mumujj~(right) channels after applying the pre-selection cuts.  The signal model assumes $\beta$\ = 1.0.  The last bin includes overflows. The ratio of the number of data events to the number of background events (and its statistical uncertainty) is also shown.  The hashed bands represent all sources of statistical and systematic uncertainty on the background prediction.}
  \label{fig:mlq}
\end{figure}

%
The distributions of the kinematic variables after performing the background-only fits in the control regions, and applying the event pre-selection requirements are shown in Figs.~\ref{fig:mll}, \ref{fig:st} and \ref{fig:mlq} for the data, background estimates, and for three LQ masses of 300, 600 and 1000~\GeV\ (with $\beta= 1.0$).

\subsection{Systematic uncertainties}
\label{sec:Systematics}

The theoretical uncertainty on the NLO cross-section is taken into account for diboson, single-top, $W$+jets, and $Z\rightarrow\tau\tau$~processes.  For the two dominant backgrounds ($t\bar{t}$ and $Z/\gamma^*$+jets) the modelling uncertainties are estimated using the symmetrised deviation from unity of the ratio of data to MC events in the $t\bar{t}$ and $Z/\gamma^*$+jets control regions, which is fitted with a linear function for $\st>400$~GeV. The modelling systematic uncertainty is then applied as a function of $\st$, in the form of a weighting factor. The choice of $\st$ for such a purpose is motivated by its sensitivity to mis-modelling of the kinematics of jets and leptons. It varies in the \eejj (\mumujj) channel between 8\% (10\%) and 25\% (30\%) for the $Z/\gamma^*$+jets background and between 6\% (10\%) and 24\% (40\%) for the $t\bar{t}$\ background. It increases for signal regions targeting higher \mlq.

The jet energy scale (JES) uncertainty depends on \pt~and \eta\ and contains additional factors, which are used to correct for pile-up effects.  They are derived as a function of the number of primary vertices in the event to take into account additional $pp$ collisions in a recorded event (in-time pile-up), or as a function of the expected number of interactions per bunch-crossing to constrain past and future collisions affecting the measurement of energies in the current bunch-crossing (out-of-time pile-up).  An additional uncertainty on the jet energy resolution (JER) is taken into account.  The relative impact on the background event yields from the JES (JER) uncertainty is between 8\% (1\%) in SR1 and 26\% (1\%) in SR8. The signal selection efficiency change due to the JES uncertainties ranges between 3\% in SR1 and 1\% in SR8, while the effect of the JER is negligible.

The electron energy scale and resolution are corrected to provide better agreement between MC predictions and data.  The uncertainties on these corrections are propagated through the analysis as sources of systematic uncertainty.  Uncertainties are taken into account for the electron trigger ($\sim0.1\%$), identification ($\sim1\%$) and reconstruction ($\sim1\%$) efficiencies, and for uncertainties associated with the isolation requirements ($\sim0.1\%$).  

Scaling and smearing corrections are applied to the \pt~of the muons in order to minimise the differences in resolution between data and MC simulated events. The uncertainty on these corrections is below 1\%.  Differences in the identification efficiency and in the efficiency of the trigger selection are taken into account and are less than 1\%.

QCD renormalisation and factorisation scales are varied by a factor of two to estimate the impact of higher orders on the signal production cross-section.  The variation is found to be approximately 14\% for all mass points.  The uncertainty on the signal cross-section related to the choice of PDF set is evaluated as the envelope of the prediction of 40 different \texttt{CTEQ6.6}~NLO error sets~\cite{Kramer:2004df}.  The uncertainty ranges from 18\% at $m_{\mathrm{LQ}}$\ = 300 \GeV\ to 56\% at $m_{\mathrm{LQ}}$\ = 1300 \GeV. These uncertainties are the same for all LQ generations.  The effect on the choice of PDF set on the signal acceptance times reconstruction efficiency is estimated using the Hessian method  \cite{Hessian}.  The final PDF uncertainties on the signal samples are approximately 1\% for most mass points, rising to 4\% for some higher LQ masses.
The impact of the choice of PDF set on the acceptance times reconstruction efficiency for each background process is estimated using the Hessian method (using the same method as for signals). The uncertainties range from 4\% in the low-mass signal regions to 17\% in the high-mass signal regions.

\subsection{Results}
\label{secl:FResults:StatAnalysis}

The observed and expected yields in three representative signal regions for the \eejj and the \mumujj channels after the combined maximum likelihood fits are shown in Table~\ref{tab:yieldsCReejj} and Table \ref{tab:yieldsCRmmjj}, respectively. The fit maximizes the likelihood constructed using the two CRs and the SR under study. When contructing the likelihood, the signal stregth and the background scale factors are treated as free parameters, the systematic uncertainties are treated as nuisance parameters. 

\begin{table}[htbp]
\footnotesize
  \begin{center}
    \begin{tabular}{l|c|c|c} \hline
      \multirow{2}{*}{\bf Yields}     & \multicolumn{3}{c}{\bf \eejj channel}                    \\ \cline{2-4}   
                                      & \textbf{SR1}             & \textbf{SR6}    & \textbf{SR8}  \\ \hline
      Observed                        & $627$                    & $8$             & $1$           \\ \hline
      Total SM                        & $(6.4\pm0.4)\times10^2$  & $11\pm2$        & $1.5\pm0.4$   \\ \hline
      $Z/\gamma^*\rightarrow\ell\ell$ & $(3.2\pm0.4)\times10^2$  & $7\pm2$         & $1.3\pm0.4$   \\
      $Z\rightarrow\tau\tau$          & $2.1\pm0.3$              & $<0.01$         & $<0.01$       \\
      $t\bar{t}$                      & $(2.4\pm0.2)\times10^2$  & $2.3\pm0.5$     & $0.12\pm0.04$ \\
      Single top                      & $19\pm3$                 & $<0.01$         & $<0.01$       \\
      Diboson                         & $22\pm3$                 & $0.8\pm0.3$     & $<0.01$       \\
      \bigcell{l}{Fake leptons
        \\(including $W$+jets)}       & $34\pm6$                 & $0.410\pm0.010$ & $0.033\pm0.006$           \\ \hline
      $\mlq$~=~300~GeV                & $(17.6\pm0.9)\times10^3$ & --              & --            \\
      $\mlq$~=~600~GeV                & --                       & $231\pm13$      & --            \\
      $\mlq$~=~1000~GeV               & --                       & --              & $5.2\pm0.3$   \\ \hline
    \end{tabular}
    \caption{Background and signal yields in three representative signal regions for LQs with masses $\mlq=$~300, 600 and 1000~\GeV\ for the \eejj\ channel (assuming $\beta$\ = 1.0). The observed number of events is also shown.  Statistical and systematic uncertainties are given.}
    \label{tab:yieldsCReejj}
  \end{center}
\end{table}

\begin{table}[htbp]
\footnotesize
  \begin{center}
    \begin{tabular}{l|c|c|c} \hline
      \multirow{2}{*}{\bf Yields}     & \multicolumn{3}{c}{\bf \mumujj channel}                      \\ \cline{2-4}    
                                        & \textbf{SR1}             & \textbf{SR6}  & \textbf{SR8}    \\ \hline
      Observed                          & 426                      & 5             & 1               \\ \hline
      Total SM                          & $(4.1\pm0.3)\times10^2$  & $7.0\pm1.2$   & $1.3\pm0.4$     \\ \hline
      $Z/\gamma^*\rightarrow\ell\ell$   & $209 \pm 18$             & $4.6 \pm 1.0$ & $0.9 \pm 0.3$   \\
      $Z\rightarrow\tau\tau$            & $0.9 \pm 0.1$            & $<0.01$       & $<0.01$         \\
      $t\bar{t}$                        & $172 \pm 18$             & $1.7 \pm 0.6$ & $0.18 \pm 0.11$ \\
      Single top                        & $14 \pm 5$               & $0.3 \pm 0.4$ & $<0.01$         \\
      Diboson                           & $14 \pm 2$               & $0.5 \pm 0.2$ & $0.19 \pm 0.05$ \\
      \bigcell{l}{Fake leptons
        \\(including $W$+jets)}         & $<0.01$                  & $<0.01$       & $<0.01$         \\ \hline
      $\mlq$~=~300~GeV                  & $(12.0\pm0.6)\times10^3$ &     --        &   --            \\
      $\mlq$~=~600~GeV                  &      --                  & $152\pm18$    &  --             \\
      $\mlq$~=~1000~GeV                 &      --                  &     --        & $3.4\pm1.3$     \\ \hline
    \end{tabular}
    \caption{Background and signal yields in three representative signal regions for LQs with masses $\mlq=$~300, 600 and 1000~\GeV\ for the \mumujj\ channel (assuming $\beta$\ = 1.0). The observed number of events is also shown.  Statistical and systematic uncertainties are given.}
    \label{tab:yieldsCRmmjj}
  \end{center}
\end{table}

No significant excess above the SM expectation is observed in any of the signal regions and a modified frequentist ${\mathrm{CL}}_s$ method \cite{0954-3899-28-10-313} is used to set limits on the strength of the LQ signal, by constructing a profile likelihood ratio. Pseudo-experiments are used to determine the limits.

\begin{figure}[htbp]
  \centering
  \subfloat[]{
    \includegraphics[width=0.5\textwidth]{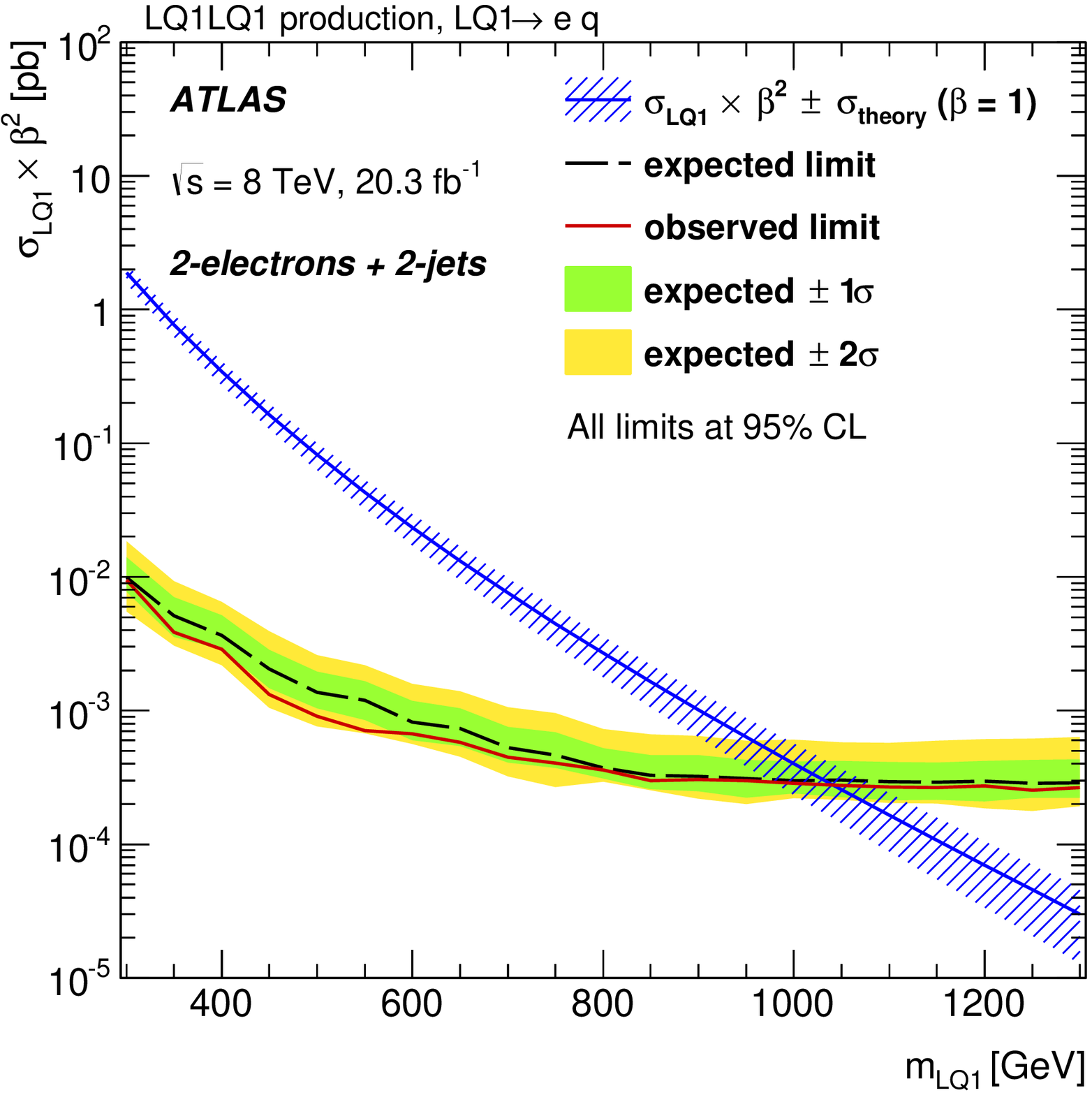}
    \label{subfig::elExclusionXSec}
  }
  \subfloat[]{
    \includegraphics[width= 0.5\textwidth]{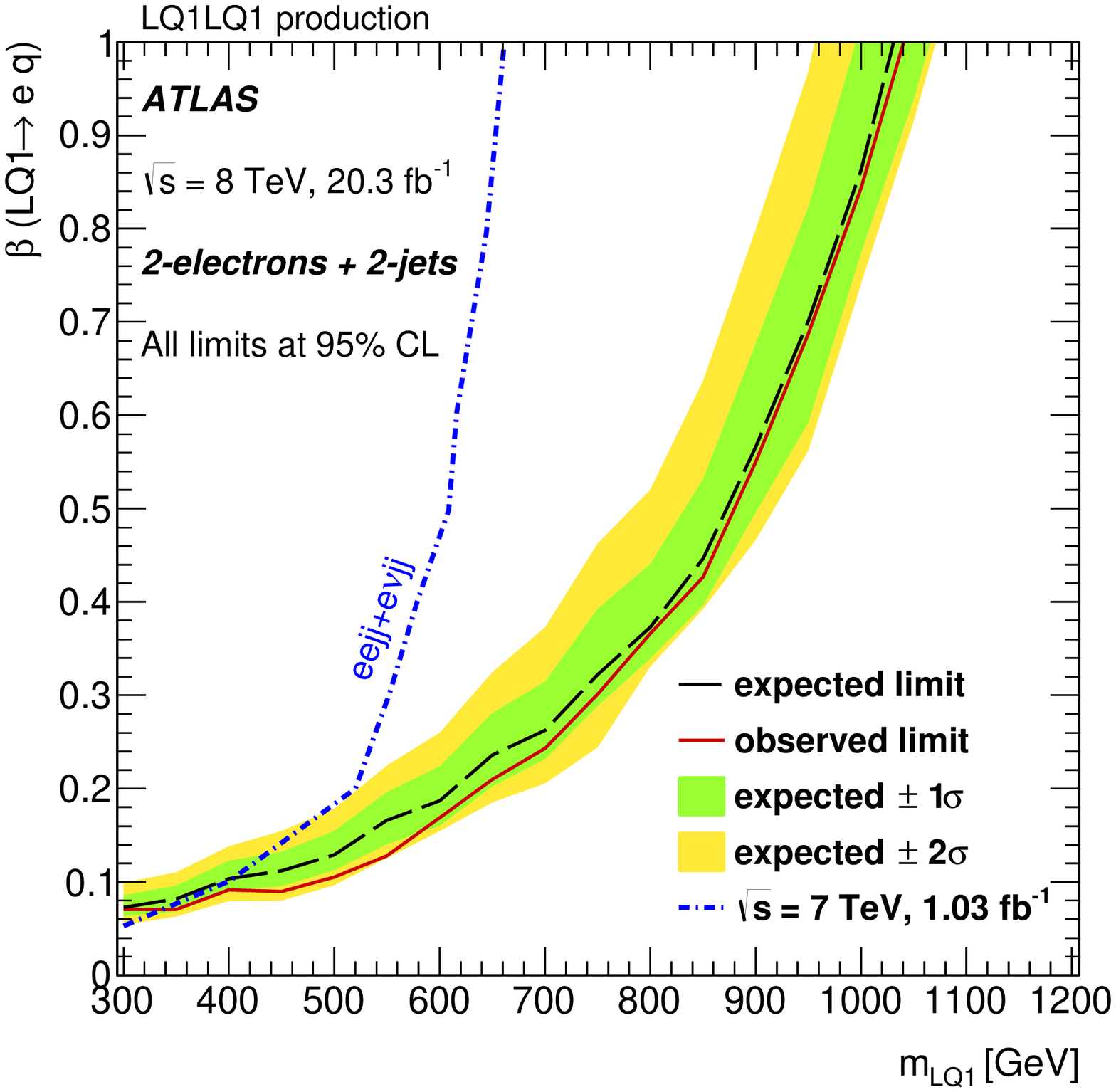}
    \label{subfig::elExclusionBeta}
  }
  \\
  \subfloat[]{
    \includegraphics[width=0.5\textwidth]{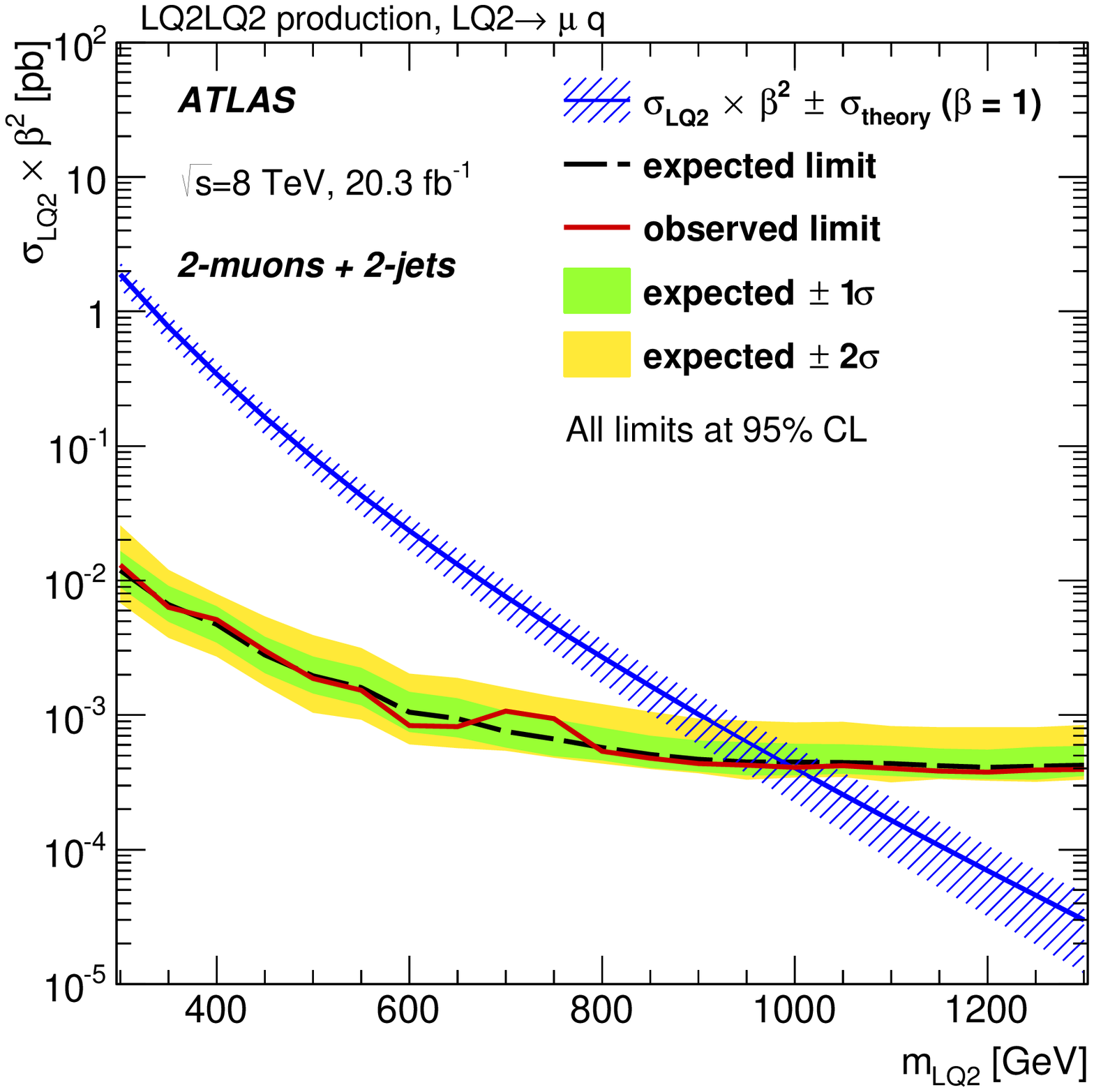}
    \label{subfig:muExclusionXSec}
  }
  \subfloat[]{
    \includegraphics[width= 0.5\textwidth]{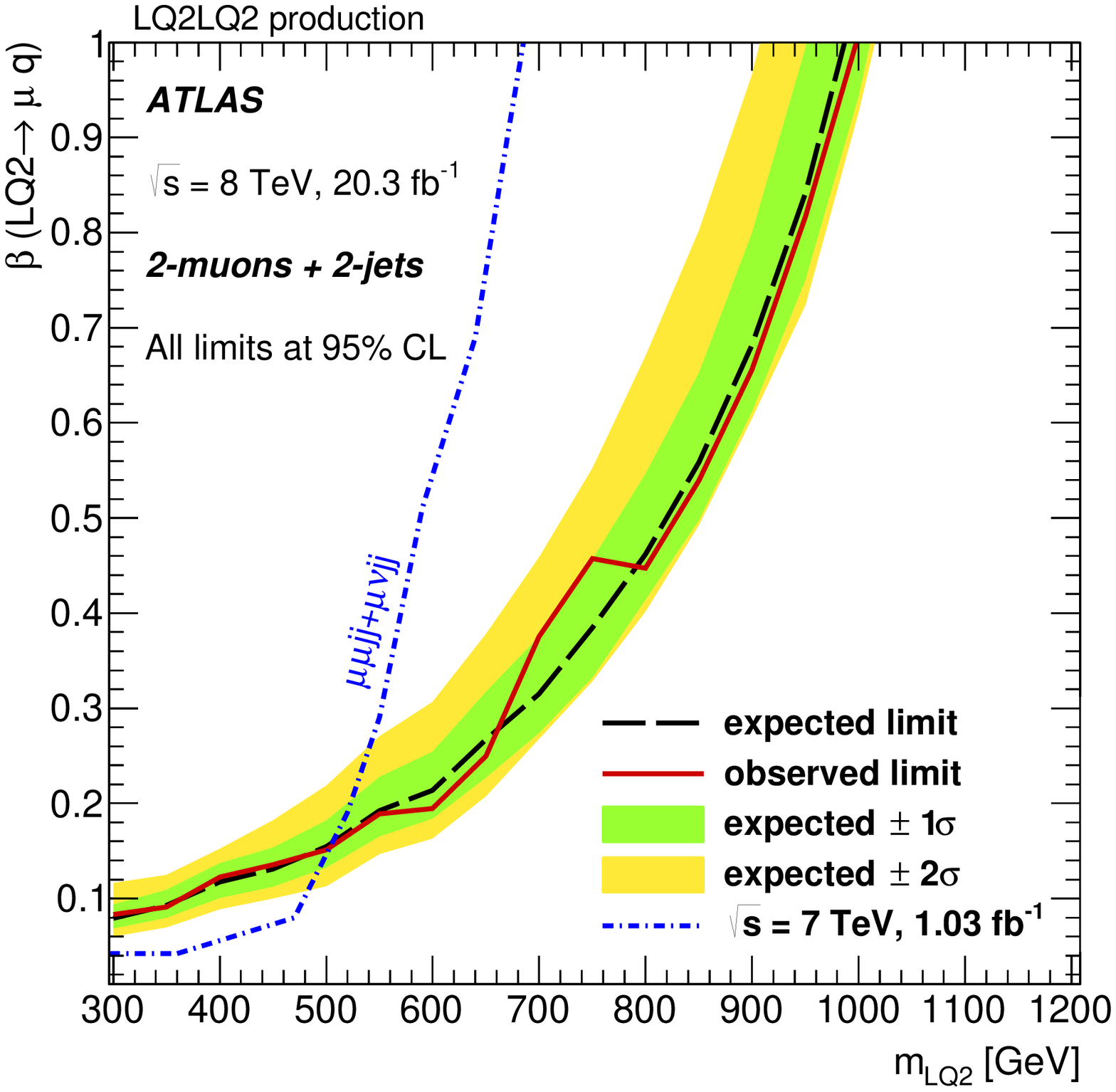}
    \label{subfig:muExclusionBeta}
  }
  \caption{The cross-section limits on scalar LQ pair-production times the square of the branching ratio as a function of mass (left) and the excluded branching ratio as a function of the LQ mass (right) to $eq$\ for the \eejj channel (top) and to $\mu q$\ for the \mumujj channel (bottom).  The $\pm1 (2) \sigma$ uncertainty bands on the expected limit represent all sources of systematic and statistical uncertainty.  The expected NLO production cross-section ($\beta = 1.0$) for scalar LQ pair-production and its corresponding theoretical uncertainty due to the choice of PDF set and renormalisation/factorisation scale are also included. The exclusion limits on $\LQone$~\cite{Aad:2011ch} and $\LQtwo$~\cite{ATLAS:2012aq} set by ATLAS in the \eejj+\enujj and \mumujj+\munujj\ search channels using 1.03~fb$^{-1}$ of data collected at $\sqrt{s}=7$~TeV are also shown. }
  \label{fig:Exclusion}
\end{figure}

The cross-section limits on scalar LQ pair-production are presented as a function of $\beta$~for both channels in Fig.~\ref{fig:Exclusion}.  Also shown are the results of the ATLAS searches for first- and second-generation LQs using 1.03 \invfb~data at \sqrts = 7 \TeV\ which also included searches in the \enujj\ and \munujj\ decay channels and therefore provide better sensitivity at low values of $\beta$.
First (second)-generation scalar LQs are excluded for $\beta=1$ at 95\% CL for $\mlqone < $~1050~GeV ($\mlqtwo <$~1000~GeV). The expected exclusion ranges are the same as the observed ones. First (second)-generation scalar LQs are excluded for $\mlqone<650$~GeV ($\mlqtwo<650$~GeV) at $\beta=0.2$ and $\mlqone<900$~GeV ($\mlqtwo<850$~GeV) at $\beta=0.5$.

%% file: Sbottom.tex
\section{Search for third-generation LQs in the $\boldsymbol{\bbMET}$~channel}

The ATLAS search for pair-production of third-generation supersymmetric partners of bottom quarks (sbottom, \sbottom) \cite{ATLAS_sbottom} is reinterpreted in terms of the LQ model, in the case where each LQ decays to a $b$-quark and a $\nu_{\tau}$ neutrino.  In the original analysis, the \sbottom~is assumed to decay via \sbottom$\rightarrow b \tilde{\chi}^{0}$, and \stop~via \stop$\rightarrow b \tilde{\chi}^{\pm}$ in the case where $m_{\tilde{\chi}^{\pm}} - m_{\tilde{\chi}^{0}}$\ is small and the $\tilde{\chi}^{\pm}$\ decay products are undetectable.  The search is performed for final states with large missing transverse momentum ($\vec{p}_{\mathrm{T}}^{\mathrm{miss}}$, with magnitude \MET) and two jets identified as originating from $b$-quarks.  The full analysis strategy is covered in Ref.~\cite{ATLAS_sbottom}.  A complete description of the analysis, including treatment of systematic uncertainties on background processes can be found there, but the event selection and background estimation methods used are summarised here for clarity.

\subsection{Object and event selection}

Events are required to have exactly two $b$-tagged \cite{MV1}\ jets with \pt\,$>$ 20\,\GeV~and \abseta\,$<$\,2.5, and \MET\ $>$\ 150 \GeV.  Additional jets in the event are accepted if they have \pt $>$ 20\,\GeV\ and \abseta\,$<$\,4.9.  Events with one or more electrons (muons) with \pT~$>$ 7 (6) \GeV\ are vetoed.  Candidate signal events are selected from data using a \met~trigger which is 99\% efficient for events passing the offline selection.  Several variables are defined and used to optimise the event selection:

\begin{itemize}
\item[--] $\dphimin$: The minimum azimuthal distance ($\Delta \phi$) between any of the leading three jets and the $\ptmissvec$.
\item[--] \meff: The scalar sum of the $\pt$~of the leading two or three jets (depending on the signal region) and the $\met$.
\item[--] \htthree: The scalar sum of the $\pt$~of all but the leading three jets.
\item[--] \mbb: The invariant mass of the two \btagged\ jets in the event.
\item[--] \mct: The contransverse mass \cite{mct_1}, used to measure the masses of pair-produced heavy particles that decay semi-invisibly (i.e. decays where one of the decay products can be detected, but the other cannot). 
\end{itemize}

In the original analysis, different signal regions were optimised according to the masses of the third-generation squark and the lightest supersymmetric particle (LSP).  In the case of the LQ model reinterpretation the signal regions corresponding to the case where the mass of the LSP is approximately zero have best sensitivity, but all the signal regions are retained for coherence with the original analysis.  The different signal region definitions are given in Table~\ref{tab:sbottom_srdef}.  Signal region A (SRA) has five different \mct~thresholds.  Signal region B (SRB) is optimised towards the region where the squark and LSP masses are approximately equal.  The signal region with the best expected limit is used for each point in the exclusion plots.

\begin{table}[hpt!]
\footnotesize
\begin{center}\renewcommand\arraystretch{1.6}
\begin{tabular}{l|c|c}
\hline
\multirow{2}{*}{Description}            & \multicolumn{2}{c}{Signal Regions}                                         \\ \cline{2-3}
                                        & SRA                                     & SRB                               \\ \hline 
Event cleaning                          & \multicolumn{2}{c}{Common to all SR}                                       \\ \hline
Lepton veto                             & \multicolumn{2}{c}{No $e$/$\mu$~after overlap removal with $\pt$~$>$~7(6) GeV for $e$($\mu$)}\\ \hline
\met                                    & $> 150$\,\GeV                           & $>250$\,\GeV                       \\ \hline
Leading jet $\pt$                       & $>$ 130\,\GeV                           & $> 150$\,\GeV                      \\ \hline
Second jet $\pt $                       & $>$ 50\,\GeV                            & $> 30$\,\GeV                       \\ \hline
Third jet $\pt $                        & veto if $>$ 50\,\GeV                    & $> 30$\,\GeV                       \\ \hline
$\Delta \phi (\ptmissvec,\mathrm{lead\,jet}$)  & -                & $> 2.5$                            \\ \hline
$b$-tagging                             & leading 2 jets                          & 2nd- and 3rd-leading jets          \\
                                        & ($ \pt >$ 50\,\GeV, $|\eta| <$ 2.5)     & ($\pt > 30$\,\GeV,$|\eta| <$ 2.5)  \\ \cline{2-3}
                                        & \multicolumn{2}{c}{$n_{b\mathchar`-\mathrm{jets}} = 2$}                         \\ \hline
$\dphimin$                              & $>$ 0.4                                 & $>$ 0.4                            \\ \hline
\bigcell{c}{\met/\meff \\ ($\meff=\met+\pt^{j_1}+\pt^{j_2}$)}           & $\met/\meff>$ 0.25                      &  --             \\ \hline
\bigcell{c}{\met/\meff \\ ($\meff=\met+\pt^{j_1}+\pt^{j_2}+\pt^{j_3}$)}  & --                      &  $\met/\meff>$ 0.25             \\ \hline
$ \mct $                                & $>$ 150, 200, 250, 300, 350\,\GeV       & -                                  \\ \hline
\htthree                                & -                                       & $< 50\,\GeV$                       \\ \hline
\mbb                                    & $>200 \gev$                             & -                                  \\ \hline
\end{tabular}
\caption{Summary of the event selection in each signal region for the \bbMET\ channel \cite{ATLAS_sbottom}.}
\label{tab:sbottom_srdef}
\end{center}
\vspace{-0.5cm}
\end{table}

\subsection{Background estimation}

The dominant background process is the production of $Z$ bosons in association with heavy-flavour jets where the $Z$~boson subsequently decays to two neutrinos ($Z(\rightarrow\nu\nu)+\bbbar$).  Its contribution is estimated from data in an opposite-sign dilepton control region.  Top quark pair-production (\ttbar) and $W$ bosons produced in association with heavy flavour quarks also contribute significantly and are normalised in dedicated control regions before being extrapolated to the signal regions using MC simulation.  Different control regions are defined for each signal region, requiring one or two leptons plus additional requirements similar to the corresponding signal region.  The contributions from $Z$+jets, $W$+jets, and top quark production are estimated simultaneously with a profile likelihood fit to the three control regions.  Contributions from diboson and \ttbar+$W/Z$ processes are estimated from MC simulation in all regions.  The contribution from multi-jet events is estimated from data by taking well-measured multi-jet events from data and smearing the jets with jet response functions taken from MC simulation and validated in data.  This procedure is described in detail in Ref. \cite{sbottom_background}.  The contribution from multi-jet events in signal regions is found to be negligible.

\subsection{Results}
\label{subsec:sbottom_results}

The number of data events observed in each signal region is reported in Table~\ref{tab:sbottom_nevents}, together with the SM background expectation after the background-only fit, and the expected number of signal events for different LQ masses.  The signal acceptance efficiency is around 2\% for all but the lowest LQ masses targeted (dropping to 0.27\% efficiency for \mlq~= 200 \GeV).  All sources of systematic and statistical uncertainty are taken into account.  The dominant systematic uncertainties on the background prediction are the jet energy scale (JES: 1--5\%) and resolution (JER: 1--8\%), and the $b$-tagging uncertainty (2--10\%).  Detector-related systematic uncertainties on the signal prediction are dominated by uncertainties on the $b$-tagging efficiency ($\sim$30\%).  The second-largest source of uncertainty is due to the JES and is around 3\%. 

The uncertainties on the signal production cross-section are estimated using the methods described in Section~\ref{sec:Systematics}.  These uncertainties are the same for all LQ genertions but the uncertainty due to the choice of PDF set varies with \mlq.  Since the third-generation analyses consider a different mass range to the first- and second-generation analyses, in this case the uncertainty due to the choice of PDF set ranges from 7.1\% at $m_{\mathrm{LQ}}$\ = 200 \GeV\ to 30\% at $m_{\mathrm{LQ}}$\ = 800 \GeV.  Effects on the acceptance due to the choice of PDF set are negligible.  

\begin{table}
\footnotesize
\begin{center}
\begin{tabular}{l|c|c|c|c|c|c}
\hline
                & \multicolumn{5}{c|}{{\bf \SRMCT, \mct $>$}}    & \bf \SRISR \\ 
                & $150$ GeV             & $200$ GeV              & $250$ GeV            &  $300$ GeV            &  $350$ GeV           &   \\ \hline
Observed        & 102                   & 48                     & 14                   & 7                     & 3                    & 65 \\ \hline
Total SM        & $94\pm13$             & $39\pm 6$              & $16\pm3$         & $5.9\pm1.1$           & $2.5\pm0.6$          & $64 \pm 10$ \\ \hline
Top quark       & $11.1 \pm 1.8$        & $2.4 \pm 1.4$          & $0.4 \pm 0.3$      & $<0.01$               & $<0.01$              & $41\pm 7$  \\
$Z$ production  &  $66 \pm 11$          & $28 \pm 5$             & $11 \pm 2$       & $4.7 \pm 0.9$         & $1.9 \pm 0.4$        & $13\pm 4$  \\
$W$ production  & $13 \pm 6$            & $5 \pm 3$          & $2.1 \pm 1.1$        & $1.0 \pm 0.5$         & $0.5 \pm 0.3$      & $8\pm 5$  \\
Others          & $4.3 \pm 1.5$         & $3.4 \pm 1.3$          & $1.8 \pm 0.6$        & $0.12 \pm 0.11$       & $0.10_{-0.10}^{+0.12}$ & $2.0\pm 1.0$  \\
Multi-jet        & $0.2 \pm 0.2$       & $0.06 \pm 0.06$        & $0.02 \pm 0.02$      & $<0.01$               & $<0.01$              & $0.16\pm 0.16$  \\\hline
$m_{\mathrm{LQ}} = 300$ GeV & $(8.5\pm 0.2)\times10^2$ & $435\pm 17$ & $96\pm 8$ & $7\pm 2$  & $0.6\pm 0.6$ & $68\pm 7$\\
$m_{\mathrm{LQ}} = 600$ GeV & $21.9\pm 0.4$   & $19.0\pm 0.4$   & $15.6\pm 0.4$ & $12.0\pm 0.3$ & $8.7\pm 0.3$ & $1.8\pm 0.1$ \\ \hline
\end{tabular}
\caption{For each signal region in the \bbMET\ channel, the observed event yield is compared with the background prediction obtained from the fit. Signal yields for different values of $m_{\mathrm{LQ}}$\ (assuming $\beta$\ = 0.0) are given for comparison.  The category `Others' includes the diboson and \ttbar+$W/Z$ processes.  Statistical, detector-related and theoretical systematic uncertainties are included, taking into account correlations \cite{ATLAS_sbottom}. 
\label{tab:sbottom_nevents}}
\end{center}
\vspace{-0.5cm}
\end{table}

\begin{figure}[htbp]
\centering 
\includegraphics[width=0.5\textwidth]{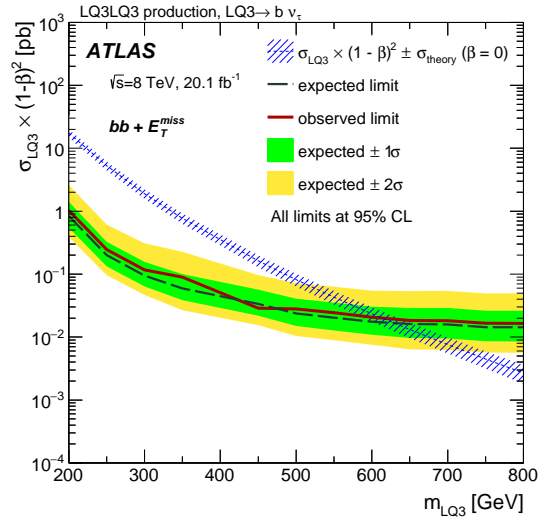}
\caption{ The expected (dashed) and observed (solid) 95\% CL upper limits on third-generation scalar LQ pair-production cross-section times the square of the branching ratio to $b\nu_{\tau}$\ as a function of LQ mass, for the \bbMET~channel.  The $\pm1 (2) \sigma$~uncertainty bands on the expected limit represent all sources of systematic and statistical uncertainty.  The expected NLO production cross-section ($\beta$\ = 0.0) for scalar LQ pair-production and its corresponding theoretical uncertainty due to the choice of PDF set and renormalisation/factorisation scale are also included.}
\label{fig:sbottom_limit}
\end{figure}

No significant excess above the SM expectation is observed in any of the signal regions. Figure~\ref{fig:sbottom_limit} shows the observed and expected exclusion limits for the scalar \LQthree~pair-production scenario obtained by taking, for each signal mass configuration, the signal region with the best expected limit. These limits are obtained using the methods described in Section~\ref{secl:FResults:StatAnalysis}.
These methods compare the observed numbers of events in the signal regions with the fitted background expectation and accounting for signal contamination in the corresponding CRs for a given model.  Pair-produced third-generation scalar LQs decaying to \bbMET~are excluded at 95\% CL for $m_{\mathrm{LQ3}}$\,$<$\,625\,\GeV.  The expected excluded range is $m_{\mathrm{LQ3}}$\,$<$\,640\,\GeV.

%% file: Stop.tex
\section{Search for third-generation LQs in the $\boldsymbol{\ttMET}$~channel}

The ATLAS search for pair-production of the supersymmetric partner of the top quark (stop quark, \stop) \cite{ATLAS_stop}~is reinterpreted in terms of the LQ model, in the case where each LQ decays to a top quark and a  $\nu_{\tau}$ neutrino.  The original analysis has dedicated signal regions targeting \stop\ decays into $t \tilde{\chi}^{0}$~and the subsequent semileptonic decay of the \ttbar\ pair.  Events compatible with \ttbar~plus extra \met\ are selected with final states containing one isolated lepton, jets, and \met.  A complete description of the analysis strategy, including the treatment of systematic uncertainties on background processes can be found in Ref.~\cite{ATLAS_stop}. The event selection and the background estimation methods are summarised here for clarity.

\subsection{Object and event selection}

Events are required to contain exactly one electron with \pt~$>$~25\,\GeV\ and \abseta\,$<$\,2.47, or muon with \pt~$>$~25\,\GeV\ and \abseta\,$<$\,2.4.  Events containing more than one electron or muon with looser identification and \pt\ requirements (10 \GeV\ for both) are vetoed. In some signal regions, events are vetoed if they are consistent with containing a hadronically decaying $\tau$ lepton.  Events are required to have a minimum of four jets with \pt~$>$~20\,\GeV\ and \abseta\,$<$\,2.5, with at least one of these passing $b$-tagging requirements \cite{MV1}.  In addition, selected events must have \met~$>$~100\,\GeV.  Several variables are used to further select signal events and reject background processes:

\begin{itemize}
\item[--] \mt: The transverse mass of the electron or muon and the \met.
\item[--] \amt~and \mttau: These are two variants on the stransverse mass (\mttwo)~\cite{mt2, stopTheTop, moreMT2} which is a generalisation of the transverse mass when applied to signatures with two invisible particles in the final state.  The asymmetric stransverse mass \amt, aims to reject dileptonic \ttbar~events where one of the leptons is not reconstructed or is outside the acceptance (and therefore adds to the \met\ of the event).  The second implementation of this variable, the $\tau$ stransverse mass $\mttwo^\tau$, targets \ttbar~events where one top decays leptonically and the other top decays into a $\tau$ that subsequently decays hadronically.  
\item[--] \topness: This variable is designed to reject dileptonic \ttbar~events where one lepton is assumed to be lost, as detailed in Ref.~\cite{topness}.  The \topness~variable is based on the minimisation of a $\chi^{2}$-type function.
\item[--] \mhadtop:  This quantity is used to reject dileptonic \ttbar~events but retain signal events that contain a hadronically decaying on-shell top quark, as in the $\LQ\rightarrow t + \nu_{\tau}$~and \topLSP~scenarios.
\item[--] $\Delta\phi(\text{jet}_{1,2}, \ptmissvec)$: The azimuthal opening angle between the leading or sub-leading jet and \ptmissvec\,used to suppress multi-jet events where \ptmissvec~is aligned with one of the leading two jets. 
\item[--] \metsig: An approximation of the \met~significance, where \HT~is defined as the scalar \pt~sum of the leading four jets.
\item[--] \HTmissSig: An object-based missing transverse momentum, divided by the per-event resolution of the jets, and shifted to the scale of the background \cite{HTmissSig}. 
\end{itemize}

The variables listed above are used to define three cut-and-count SRs and one shape-fit SR.  Table~\ref{tab:stop_eventselection}~details the event selections for these signal regions.  The SR labelled \SRtNboost~targets LQ/stop masses of $\gtrsim$700\,\GeV~and takes advantage of the `boosted' topology of such a heavy parent particle. The selection assumes that either all decay products of the hadronically decaying top quark, or at least the decay products of the hadronically decaying $W$ boson, collimate into a jet reconstructed with a radius parameter $R$ = 1.0 \cite{jetSubstructure, jetTrimming}. 

\begin{table}
\begin{center}
\setlength{\tabcolsep}{0.36pc}
\renewcommand{\arraystretch}{1.5}
{\footnotesize
\begin{tabular}{ l | l | l | l | l }
\hline 
 & \SRtNonep & \SRtNtwox & \SRtNthreep &  \SRtNboost \\
\hline 
Lepton & \multicolumn{4}{ |c }{ $=1$ lepton } \\
\hline 
Jets & $\ge 4$ with $\pT>$ & $\ge 4$ with $\pT>$ &  $\ge 4$ with $\pT>$ & $\ge 4$ with $\pT>$ \\
                    & $60,60,40,25$\,\GeV & $80,60,40,25$\,\GeV &  $100,80,40,25$\,\GeV &  $75,65,40,25$\,\GeV \\
\hline
$b$-tagging & \multicolumn{4}{ |c }{$\ge 1 b$-tag (70\% eff.) amongst four selected jets}\\
\hline
Large-$R$ jet & \multicolumn{3}{ |c| }{--} & $\ge 1$, $\pT>270$\,\GeV\\
  & \multicolumn{3}{ |c| }{ } & and $m > 75$\,\GeV\\
\hline
$\Delta\phi(\text{jet}^{\mathrm{large}R}_2, \ptmissvec )$ & \multicolumn{3}{ |c| }{--} & $> 0.85$ \\
\hline
$\met$ & $>100$\,\GeV & $>200$\,\GeV & $>320$\,\GeV & $>315$\,\GeV \\
\hline
$\mt$ & $>60$\,\GeV & $>140$\,\GeV &  $>200$\,\GeV &  $>175$\,\GeV \\
\hline
$\amt$ & -- & $>170$\,\GeV &  $>170$\,\GeV &  $>145$\,\GeV \\
\hline
$\mttau$ & -- & -- &  $>120$\,\GeV & -- \\
\hline
\textit{Topness} & -- & -- &  -- &  $>7$ \\
\hline
$\mtophad$ & $\in$ [$130$, $205$]\,\GeV & $\in$ [$130$, $195$]\,\GeV &  $\in$ [$130$, $250$]\,\GeV &  \\
\hline
$\tau$-veto &  loose $\tau$ particle ID & -- &  -- &  modified, see \cite{ATLAS_stop}. \\
\hline
$\Delta R(b\rm{-jet}, \ell)$    & $< 2.5$ & -- & $< 3$ &$< 2.6$ \\
\hline
$\metsig$ & $> 5\,\GeV^{1/2}$ & \multicolumn{3}{ |c }{--} \\ 
\hline
$\HTmissSig$ & -- & \multicolumn{2}{ |c| }{$>12.5$} & $>10$ \\
\hline
$\Delta\phi(\text{jet}_i, \vec{p}_\text{T}^\text{miss})$ & $>0.8$ {\scriptsize ($i=1,2$)} & $>0.8$ {\scriptsize($i=2$)} & -- & $>0.5,0.3$ {\scriptsize($i=1,2$)} \\ 
\hline
\hline
Exclusion setup & shape-fit in \mt\ and& \multicolumn{3}{ |c }{cut-and-count}   \\
 & \met. & \multicolumn{3}{ |c }{ }\\
\hline
\end{tabular}
}
\end{center}

\caption{Selection criteria for the four SRs (\SRtNonep, \SRtNtwox, \SRtNthreep, and \SRtNboost) employed to search for $\LQLQ\rightarrow\ttMET$~events \cite{ATLAS_stop}. The details of the limit-setting procedure for the exclusion setup can be found in Section~\ref{sec:stop_results}.
\label{tab:stop_eventselection}
}
\end{table}

\subsection{Background estimation}

The dominant sources of background are the production of \ttbar~events and $W$+jets where the $W$\ boson decays leptonically. Other background processes considered are single top, dibosons, $Z$+jets, \ttbar~produced in association with a vector boson (\ttbar$V$), and multi-jets.

The predicted numbers of \ttbar~and $W$+jets background events in the SRs are estimated from data using a fit to the number of observed events in dedicated control regions.  Each SR has an associated CR for each of the \ttbar\ and $W$+jets backgrounds.  The CRs are designed to select events as similar as possible to those selected by the corresponding SR while keeping the contamination from other backgrounds and potential signal low.  This is achieved by e.g. requiring that 60 $<$~\mt\ $<$~90 \GeV\ and in the case of the $W$+jets CR, inverting the $b$-jet requirement so that it becomes a $b$-jet veto.  The simulation is used to extrapolate the background predictions into the signal region. The background fit predictions are validated using dedicated event samples, referred to as validation regions (VRs), and one or more VR is defined for each of these. Most VRs are defined by changing the \mt\ windows to 90 $<$~\mt\ $<$~120 \GeV.  The VRs are designed to be kinematically close to the associated SRs to test the background estimates in regions of phase space as similar as possible to the SRs. 

The multi-jet background is estimated from data using a matrix method described in Refs.~\cite{stop_matrix1,stop_matrix2}.  The contribution is found to be negligible.  All other (small) backgrounds are determined entirely from simulation and normalised to the most accurate theoretical cross-sections available.

\subsection{Results}
\label{sec:stop_results}

The number of events observed in each signal region is reported in Tables~\ref{tab:stop_nevents_count}~and \ref{tab:stop_nevents_shape}, together with the SM background expectation and the expected number of signal events for different LQ masses.  The signal acceptance is between 1.5\% and 3\% depending on the LQ mass.  All sources of systematic uncertainty and statistical uncertainty are taken into account.  The dominant sources of uncertainty on the background prediction come from uncertainties related to the JES, JER, \ttbar~background modelling, the $b$-tagging efficiency, and statistical uncertainties.  

\begin{table}[ht]
\begin{center}
\footnotesize
\setlength{\tabcolsep}{0.2pc}
\begin{tabular}{l|c|c|c}
\hline 
                              & \SRtNtwox          & \SRtNthreep           & \SRtNboost        \\ \hline
Observed                      & 12                 & 5                     & 5                 \\ \hline
Total SM                      & 13 $\pm$ 2     & 5.0 $\pm$ 0.9         & 3.3 $\pm$ 0.7     \\ \hline
\ttbar                        & 6.5 $\pm$ 1.7      & 2.0 $\pm$ 0.6         & 1.1 $\pm$ 0.4     \\
$W$+jets                      & 2.1 $\pm$ 0.5      & 0.9 $\pm$ 0.3       & 0.28 $\pm$ 0.14   \\
Single top                    & 1.1 $\pm$ 0.5      & 0.54 $\pm$ 0.19       & 0.39 $\pm$ 0.15   \\
Diboson                      & 1.4 $\pm$ 0.6      & 0.9 $\pm$ 0.3       & 0.7 $\pm$ 0.3   \\
$Z$+jets                      & 0.009 $\pm$ 0.005  & 0.003 $\pm$ 0.002     & 0.004 $\pm$ 0.002 \\
\ttbar$V$                     & 2.0 $\pm$ 0.6      & 0.8 $\pm$ 0.3       & 0.9 $\pm$ 0.3   \\ \hline
$m_{\mathrm{LQ}}$ = 300 \GeV    & 20 $\pm$ 3     & 3.4 $\pm$ 1.1        & 3.8  $\pm$ 1.2   \\
$m_{\mathrm{LQ}}$ = 600 \GeV    & 10.7 $\pm$ 0.3    & 7.9 $\pm$ 0.3       & 8.9  $\pm$ 0.3  \\ \hline
\end{tabular}
\caption{The number of observed events in the three cut-and-count signal regions, together with the expected number of background events and signal events for different $\LQ$\ masses (assuming $\beta$\ = 0.0) in the \ttMET\ channel \cite{ATLAS_stop}.}
\label{tab:stop_nevents_count}
\end{center}
\end{table}

\begin{table}[ht]
\begin{center}
\footnotesize
\setlength{\tabcolsep}{0.2pc}
\begin{tabular}{l|c|c|c|c} \hline 
                            &\multicolumn{4}{|c}{\SRtNonep}                                                                              \\             
                            &~$125 < \met  < 150$\,\GeV  &~$125 < \met  < 150$\,\GeV  &~$\met > 150$\,\GeV          &~$\met > 150$\,\GeV  \\ 
                            &~$120 < \mt < 140$\,\GeV    &~$\mt > 140$\,\GeV          &~$120 < \mt < 140$\,\GeV     &~$\mt > 140$\,\GeV   \\ \hline
Observed                    & 117                        & 163                        & 101                         & 217                 \\ \hline
Total SM                    & $(1.4 \pm 0.2)\times10^2$               & $(1.5 \pm 0.2)\times10^2$               & 98 $\pm$ 13                 & $(2.4 \pm 0.3)\times10^2$        \\ \hline
\ttbar                      & $(1.2 \pm 0.2)\times10^2$               & $(1.4 \pm 0.2)\times10^2$               & 85 $\pm$ 12                 & ($2.0 \pm 0.3)\times10^2$        \\
$W$+jets                    & 7 $\pm$ 3              & 6 $\pm$ 3              & 4.6 $\pm$ 1.5               & 10 $\pm$ 4          \\
Single top                  & 5 $\pm$ 2              & 6 $\pm$ 2              & 6$\pm$ 2               & 9 $\pm$ 4       \\
Diboson                    & 0.29 $\pm$ 0.18            & 0.8 $\pm$ 0.5              & 0.3 $\pm$ 0.3             & 0.30 $\pm$ 0.15     \\
$Z$+jets                    & 0.17 $\pm$ 0.08            & 0.24 $\pm$ 0.12            & 0.30 $\pm$ 0.15             & 0.5 $\pm$ 0.3     \\
\ttbar$V$                   & 1.5 $\pm$ 0.5              & 2.9 $\pm$ 0.9              & 2.5 $\pm$ 0.8               & 11 $\pm$ 3      \\ \hline
$m_{\mathrm{LQ}}$ = 300 \GeV  & 28 $\pm$ 3            & 77 $\pm$ 6            & 64 $\pm$ 5             & 269 $\pm$ 10      \\
$m_{\mathrm{LQ}}$ = 600 \GeV  & 0.15 $\pm$ 0.04         & 0.62 $\pm$ 0.08         & 0.83 $\pm$ 0.09            & 18.8 $\pm$ 0.4    \\ \hline
\end{tabular}
\caption{The number of observed events in the shape-fit signal region, together with the expected number of background events and signal events for different $\LQ$\ masses (assuming $\beta$\ = 0.0) in the \ttMET\ channel \cite{ATLAS_stop}. }
\label{tab:stop_nevents_shape}
\end{center}
\end{table}

Detector-related systematic effects are evaluated for signal using the same methods used for the backgrounds (see Ref.~\cite{ATLAS_stop}\,for details).  The dominant detector-related systematic effects are the uncertainties on the JES (4\%) and the $b$-tagging efficiency (3\%).  

The uncertainties on the signal production cross-section are estimated using the methods described in Section~\ref{sec:Systematics}.  The effect on the choice of PDF set on the signal acceptance is less than 1\% for most mass points, but increases to 1.7\% for $m_{\mathrm{LQ}}$\ = 800 \GeV.  

Similar methods as described in Section~\ref{secl:FResults:StatAnalysis} are used to assess the compatibility of the SM background-only hypothesis with the observations in the signal regions.  The observed number of events is found to agree well with the expected number of background events in all signal regions.  No significant excess over the expected background from SM processes is observed and the data are used to derive one-sided limits at 95\% CL. The results are obtained from a profile likelihood-ratio test following the ${\mathrm{CL}}_s$ prescription \cite{0954-3899-28-10-313}. The likelihood of the simultaneous fit is configured to include all CRs and one SR or shape-fit bin.  The `exclusion setup' event selection is applied (see Table~\ref{tab:stop_eventselection}), and all uncertainties except the theoretical signal uncertainty are included in the fit.  

\begin{figure}[htbp]
\centering 
\includegraphics[width=0.5\textwidth]{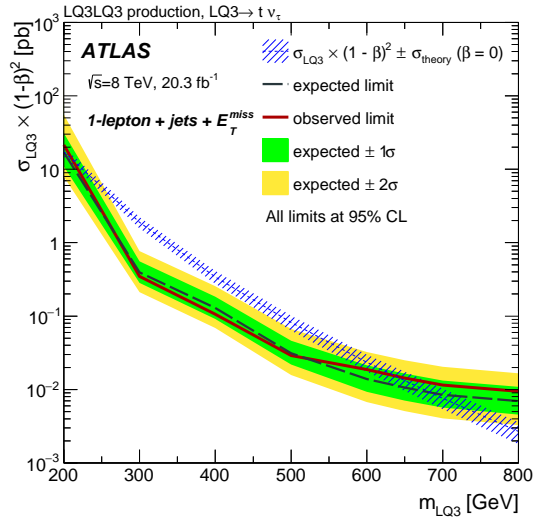}
\caption{ The expected (dashed) and observed (solid) 95\% CL upper limits on the third-generation scalar LQ pair-production cross-section times the square of the branching ratio to $t\nu_{\tau}$\ as a function of LQ mass, for the \ttMET~channel.  The $\pm1 (2) \sigma$~uncertainty bands on the expected limit represent all sources of systematic and statistical uncertainty.  The expected NLO production cross-section ($\beta$\ = 0.0) for third-generation scalar LQ pair-production and its corresponding theoretical uncertainty due to the choice of PDF set and renormalisation/factorisation scale are also included.}
\label{fig:stop_limit}
\end{figure}

Exclusion limits are obtained by selecting a priori the signal region with the lowest expected ${\mathrm{CL}}_s$ value for each signal grid point.  The expected and observed limits on the $\LQLQ\rightarrow$~\ttMET~process are shown in Fig.~\ref{fig:stop_limit}.  Third-generation scalar LQs decaying to \ttMET~are excluded at 95\% CL in the mass range 210\,$<$\,$m_{\mathrm{LQ3}}$\,$<$\,640\,\GeV.  The expected exclusion range is 200\,$<$\,$m_{\mathrm{LQ3}}$\,$<$\,685\,\GeV.  The limits for stop production in the case where the neutralino is massless are slightly stronger than the limits set on LQ3 production since the nominal stop limits consider a mostly right-handed stop.  This leads to the top quarks being polarised in such a way that the acceptance increases.  The limit worsens at low mass, due to the effect of greater contamination from top backgrounds.

%% file: Conclusions.tex
\section{Summary and conclusions}
Searches for pair-production of first-, second- and third-generation scalar leptoquarks have been performed with the ATLAS detector at the LHC using an integrated luminosity of 20~\invfb~of data from $pp$ collisions at $\sqrt{s} = 8$~\TeV.  No significant excess above the SM background expectation is observed in any channel.  The results are summarised in Table \ref{tab:limit_summary}.

\begin{table}[h!]
\footnotesize
\begin{center}
\setlength{\tabcolsep}{0.2pc}
\begin{tabular}{l|c|c}
\hline 
\multirow{2}{*}{Decay channel}  & \multicolumn{2}{c}{Excluded range (95\% CL)}                        \\
                                & Expected                         & Observed                         \\ \hline
\eejj ($\beta$ = 1.0)           & $\mlqone < 1050$~GeV             & $\mlqone < 1050$~GeV             \\ \hline
\mumujj ($\beta$ = 1.0)         & $\mlqtwo < 1000$~GeV             & $\mlqtwo < 1000$~GeV             \\ \hline
\bbMET ($\beta$ = 0.0)          & $m_{\mathrm{LQ3}} < 640$ GeV       & $m_{\mathrm{LQ3}} < 625$~GeV       \\ \hline
\ttMET ($\beta$ = 0.0)          & $200 < m_{\mathrm{LQ3}} < 685$~GeV & $210 < m_{\mathrm{LQ3}} < 640$~GeV \\ \hline
\end{tabular}
\caption{Expected and observed exclusion ranges at 95\% CL for each of the four LQ decay channels considered.}
\label{tab:limit_summary}
\end{center}
\end{table}
\FloatBarrier

The results presented here significantly extend the sensitivity in mass compared to previous searches.  Low-mass regions are also considered and limits on the cross-sections are provided for the different final states analysed.   Since $\beta$~is not constrained by the theory, searches in the low mass regions are also important in order to extract limits for low-$\beta$ values for the \LQone\ and \LQtwo\ analyses.

%% file: Acknowledgements.tex

We thank CERN for the very successful operation of the LHC, as well as the
support staff from our institutions without whom ATLAS could not be
operated efficiently.

We acknowledge the support of ANPCyT, Argentina; YerPhI, Armenia; ARC, Australia; BMWFW and FWF, Austria; ANAS, Azerbaijan; SSTC, Belarus; CNPq and FAPESP, Brazil; NSERC, NRC and CFI, Canada; CERN; CONICYT, Chile; CAS, MOST and NSFC, China; COLCIENCIAS, Colombia; MSMT CR, MPO CR and VSC CR, Czech Republic; DNRF, DNSRC and Lundbeck Foundation, Denmark; IN2P3-CNRS, CEA-DSM/IRFU, France; GNSF, Georgia; BMBF, HGF, and MPG, Germany; GSRT, Greece; RGC, Hong Kong SAR, China; ISF, I-CORE and Benoziyo Center, Israel; INFN, Italy; MEXT and JSPS, Japan; CNRST, Morocco; FOM and NWO, Netherlands; RCN, Norway; MNiSW and NCN, Poland; FCT, Portugal; MNE/IFA, Romania; MES of Russia and NRC KI, Russian Federation; JINR; MESTD, Serbia; MSSR, Slovakia; ARRS and MIZ\v{S}, Slovenia; DST/NRF, South Africa; MINECO, Spain; SRC and Wallenberg Foundation, Sweden; SERI, SNSF and Cantons of Bern and Geneva, Switzerland; MOST, Taiwan; TAEK, Turkey; STFC, United Kingdom; DOE and NSF, United States of America. In addition, individual groups and members have received support from BCKDF, the Canada Council, CANARIE, CRC, Compute Canada, FQRNT, and the Ontario Innovation Trust, Canada; EPLANET, ERC, FP7, Horizon 2020 and Marie Skłodowska-Curie Actions, European Union; Investissements d'Avenir Labex and Idex, ANR, Region Auvergne and Fondation Partager le Savoir, France; DFG and AvH Foundation, Germany; Herakleitos, Thales and Aristeia programmes co-financed by EU-ESF and the Greek NSRF; BSF, GIF and Minerva, Israel; BRF, Norway; the Royal Society and Leverhulme Trust, United Kingdom.

The crucial computing support from all WLCG partners is acknowledged
gratefully, in particular from CERN and the ATLAS Tier-1 facilities at
TRIUMF (Canada), NDGF (Denmark, Norway, Sweden), CC-IN2P3 (France),
KIT/GridKA (Germany), INFN-CNAF (Italy), NL-T1 (Netherlands), PIC (Spain),
ASGC (Taiwan), RAL (UK) and BNL (USA) and in the Tier-2 facilities
worldwide.

%% file: ScalarLQPaper.bbl
\providecommand{\href}[2]{#2}\begingroup\raggedright\endgroup

%% file: atlas_authlist.tex
\begin{flushleft}
{\Large The ATLAS Collaboration}

\bigskip

G.~Aad$^{\rm 85}$,
B.~Abbott$^{\rm 113}$,
J.~Abdallah$^{\rm 151}$,
O.~Abdinov$^{\rm 11}$,
R.~Aben$^{\rm 107}$,
M.~Abolins$^{\rm 90}$,
O.S.~AbouZeid$^{\rm 158}$,
H.~Abramowicz$^{\rm 153}$,
H.~Abreu$^{\rm 152}$,
R.~Abreu$^{\rm 116}$
% ...
\bigskip
\\
% ...
$^{11}$ Institute of Physics, Azerbaijan Academy of Sciences, Baku, Azerbaijan\\
% ...
$^{85}$ CPPM, Aix-Marseille Universit{\'e} and CNRS/IN2P3, Marseille, France\\
% ...
$^{90}$ Department of Physics and Astronomy, Michigan State University, East Lansing MI, United States of America\\
% ...
$^{107}$ Nikhef National Institute for Subatomic Physics and University of Amsterdam, Amsterdam, Netherlands\\
% ...
$^{113}$ Homer L. Dodge Department of Physics and Astronomy, University of Oklahoma, Norman OK, United States of America\\
% ...
$^{116}$ Center for High Energy Physics, University of Oregon, Eugene OR, United States of America\\
% ...
$^{151}$ Institute of Physics, Academia Sinica, Taipei, Taiwan\\
$^{152}$ Department of Physics, Technion: Israel Institute of Technology, Haifa, Israel\\
$^{153}$ Raymond and Beverly Sackler School of Physics and Astronomy, Tel Aviv University, Tel Aviv, Israel\\
% ...
$^{158}$ Department of Physics, University of Toronto, Toronto ON, Canada\\
% ...
\end{flushleft}